\documentclass{aa}

\usepackage[varg]{txfonts}
\usepackage{graphicx}
\usepackage{color} 
\usepackage{amsmath}
\usepackage{multirow} 
\bibpunct{(}{)}{;}{a}{}{,} 

\begin{document}

\title{Unveiling the physical conditions of the youngest disks:}
\subtitle{A warm embedded disk in L1527}

\author{Merel L.R. van 't Hoff\inst{1}
\and John J. Tobin\inst{1,2}
\and Daniel Harsono\inst{1}
\and Ewine F. van Dishoeck\inst{1,3}}

\institute{Leiden Observatory, Leiden University, P.O. box 9513, 2300 RA Leiden, The Netherlands
\and Homer L. Dodge Department of Physics and Astronomy, University of Oklahoma, 440 W. Brooks Street, Norman, OK 73019, USA
\and Max-Planck-Institut f\"ur Extraterrestrische Physik, Giessenbachstrasse 1, 85748 Garching, Germany }

\date{}

\abstract {Protoplanetary disks have been studied extensively, both physically and chemically, to understand the environment in which planets form. However, the first steps of planet formation are likely to occur already when the protostar and disk are still embedded in their natal envelope. The initial conditions for planet formation may thus be provided by these young embedded disks, of which the physical and chemical structure is poorly characterized.} 
{We aim to constrain the midplane temperature structure, one of the critical unknowns, of the embedded disk around L1527. In particular, we set out to determine whether there is an extended cold outer region where CO is frozen out, as is the case for Class~II disks. This will show whether young disks are indeed warmer than their more evolved counterparts, as is predicted by physical models. } 
{We use archival ALMA data of $^{13}$CO $J=2-1$, C$^{18}$O $J=2-1$ and N$_2$D$^+ J=3-2$ to directly observe the midplane of the near edge-on L1527 disk. The optically thick CO isotopologues allow us to derive a radial temperature profile for the disk midplane, while N$_2$D$^+$, which can only be abundant when CO is frozen out, provides an additional constraint on the temperature. Moreover, the effect of CO freeze-out on the $^{13}$CO, C$^{18}$O and N$_2$D$^+$ emission is investigated using 3D radiative transfer modeling.}
{Optically thick $^{13}$CO and C$^{18}$O emission is observed throughout the disk and inner envelope, while N$_2$D$^+ $ is not detected. Both CO isotopologues have brightness temperatures $\gtrsim$ 25~K along the midplane. Disk and envelope emission can be disentangled kinematically, because the largest velocities are reached in the disk.  A power law radial temperature profile constructed using the highest midplane temperature at these velocities suggest that the temperature is above 20~K out to at least 75~AU, and possibly throughout the entire 125~AU disk. The radiative transfer models show that a model without CO freeze-out in the disk matches the C$^{18}$O observations better than a model with the CO snowline at $\sim$70~AU. In addition, there is no evidence for a large (order of magnitude) depletion of CO. }
{The disk around L1527 is likely to be warm enough to have CO present in the gas phase throughout the disk, suggesting that young embedded disks can indeed be warmer than the more evolved Class~II disks.}


\maketitle


\section{Introduction}

Disks around young stars play a fundamental role in the star and planet formation process by regulating mass accretion and providing the mass reservoirs for forming planets. The physical structure of the disk is intimately related to its chemical structure, which in turn determines the initial planet composition. Many studies have therefore focused on the evolved protoplanetary disks around pre-main sequence stars, and these Class II disks are now becoming well characterized both physically \citep[e.g.,][]{Andrews2010,Schwarz2016,Ansdell2016} and chemically \citep[e.g.,][]{Dutrey1997,Thi2004,Oberg2010,Huang2017}. The thermal and chemical structure that has emerged consists of a cold dense midplane where CO is frozen out, a warm molecular layer where CO and other species are in the gas phase, and a tenuous upper layer where molecules are photodissociated by UV radiation \citep[e.g.,][]{Aikawa2002}. In addition, disk masses are typically low; average dust masses range between 5 and 15 $M_{\oplus}$ (depending on the star forming region) and gas masses are generally $\lesssim 1 M_{\rm{jup}}$ \citep{Andrews2013,Ansdell2016,Ansdell2017,Barenfeld2016,Pascucci2016,Law2017}.

However, it is now becoming clear that the first steps of planet formation likely occur at an earlier stage, when the disk is still forming within the infalling envelope (Class 0 and I sources). Grain growth begins during this embedded stage \citep{Kwon2009,Pagani2010,Foster2013,Miotello2014}, and the HL Tau images suggest that even the formation of larger bodies may have already started before the envelope has fully dissipated \citep{ALMAPartnership2015}. Thus, young embedded disks may reflect the true initial conditions for planet formation. 

\begin{table*}
\addtolength{\tabcolsep}{-2pt}
\caption{Overview of the molecular line observations toward L1527.
\label{tab:Lineparameters}} 
\centering
\begin{tabular}{l c c c c c c c c c}
    \hline\hline
    \\[-.3cm]
    
    Molecule & Transition & Frequency & $A_{\rm{ul}}$\tablefootmark{a} & $g_{\rm{up}}$\tablefootmark{b} & $E_{\rm{up}}$\tablefootmark{c} & Beam & $\Delta v$\tablefootmark{d} & rms & rms \\ 
    & & (GHz) & (s$^{-1}$) & & (K) & ($\arcsec$) & (km s$^{-1}$) & (mJy beam$^{-1}$  & (K \\
    & & & & & & & & channel$^{-1}$) & channel$^{-1}$)\\
    \hline 
    \\[-.3cm]
    $^{13}$CO & $J=2-1$  & 220.39868 & 6.038$\times$10$^{-7}$ & 5 & 15.87 & 0.32$\times$0.18 (PA = 39.0$^{\circ}$) & \hspace{0.2cm}0.08\tablefootmark{e} & \hspace{0.2cm}10.2\tablefootmark{f} & \hspace{0.2cm}4.4\tablefootmark{f} \\
    C$^{18}$O & $J=2-1$  & 219.56035 & 6.011$\times$10$^{-7}$ & 5 & 15.81 &  0.32$\times$0.19 (PA = 38.7$^{\circ}$) & \hspace{0.2cm}0.04\tablefootmark{e} & \hspace{0.4cm}8.7\tablefootmark{f} & \hspace{0.2cm}3.6\tablefootmark{f}\\
    N$_2$D$^+$ & $J=3-2$ & 231.32183 & 7.138$\times$10$^{-4}$ & 7 & 22.20 & 0.93$\times$0.71 (PA = 10.7$^{\circ}$) & 0.08 & 10.4 & 0.4 \\
    \hline
\end{tabular}
\tablefoot{\tablefoottext{a}{Einstein-A coefficient.}\tablefoottext{b}{Upper level degeneracy.}\tablefoottext{c}{Upper level energy.}\tablefoottext{d}{Velocity resolution.}\tablefoottext{e}{Velocities were binned to 0.16 km~s$^{-1}$.}\tablefoottext{f}{In 0.16 km~s$^{-1}$ channels.}}
\addtolength{\tabcolsep}{+2pt}
\end{table*}


Efforts to detect these embedded disks have been made for the past 20 years, but early studies of the dust continuum were hampered by a lack of sensitivity and/or spatial resolution \citep[e.g.,][]{Keene1990,Looney2000,Jorgensen2009}. While recent interferometric surveys have resolved disk-like structures in the dust around many protostars \citep{Tobin2015,Segura-Cox2016}, molecular line observations are necessary to establish whether these disk structures are in fact rotationally supported. Keplerian rotation has indeed been observed for a handful Class 0 sources \citep{Tobin2012,Murillo2013,Codella2014,Lindberg2014,Yen2017}, and more disks have been detected around the less embedded Class I protostars \citep[e.g.,][]{Brinch2007,Lommen2008,Lee2010,Takakuwa2012,Harsono2014}. 

So far, studies of these young disk systems have mostly focused on disk size, kinematics and disk formation mechanisms \citep[e.g.,][]{Yen2013,Ohashi2014,Harsono2014}, or the chemical structure at the disk-envelope interface \citep{Sakai2014b,Sakai2014a,Murillo2015}. Although a few studies have set out to model the disk radial density profile \citep[e.g.,][hereafter T13 and A17, resp.]{Tobin2013,Aso2017}, the physical and chemical structure of embedded disks remains poorly characterized. A simple but crucial question, for example, is whether these young disks are warm ($T\gtrsim$ 20~K, i.e. warmer than the CO freeze-out temperature) or cold (have a large region where $T\lesssim$ 20~K), as this will strongly affect disk evolution as well as the molecular composition. 

The effect of temperature on the chemical structure can be readily seen in the snowline locations of the major volatile species. The snowline is the midplane radius where the temperature drops below the freeze-out temperature of a given molecule and hence this molecule freezes out from the gas phase onto the dust grains. The sequential freeze out of molecules causes molecular abundances and elemental ratios (like the C/O ratio) in both the gas and ice to vary with radius \citep[e.g.,][]{Oberg2011}. This in turn affects the composition of planets formed at different orbital distances \citep[e.g.,][]{Madhusudhan2014,Walsh2015,Eistrup2016,Ali-Dib2017}. In addition, the solid-state formation of high abundances of complex molecules, which starts from CO ice \citep[e.g.,][]{Tielens1982,Garrod2006,Chuang2016}, will be impeded if the disk temperature is higher than the freeze-out temperature of CO ($\gtrsim$20~K). Furthermore, the location where grain growth is efficient is related to the water snowline \citep[e.g.,][]{Stevenson1988,Schoonenberg2017}. The thermal structure of the disk is thus a crucial ingredient in the planet formation process.  

Modeling studies of evolved disks without an envelope have shown that the mass accretion rate is an important parameter for the disk temperature; higher accretion rates result in warmer disks \citep{Davis2005,Garaud2007,Min2011}. Since accretion rates are expected to decrease on average as the protostar evolves, disks in the embedded phase may be warmer than their more evolved counterparts. The envelope itself also influences the disk temperature as it serves as an isolating blanket \citep{d'Alessio1997,Whitney2003}. Specifically, \citet{Harsono2015} showed that both a high accretion rate and the presence of an envelope affects the location of $T \lesssim$ 40~K dust, shifting the CO snowline to larger radii compared to disks without an envelope. 

L1527 IRS (IRAS 04368+2557) is a good protostellar source to investigate whether young embedded disks are indeed warm, or have a large cold outer region where CO is frozen out, like the more evolved Class II disks. This protostar located in the Taurus molecular cloud ($d =$140 pc, \citealt{Torres2007}) is a borderline Class 0/I object (see e.g. the discussion in T13), with a bolometric luminosity of $\sim1.9-2.6$ $L_{\sun}$ \citep{Tobin2008,Kristensen2012}. \textit{Spitzer} images revealed bright bipolar outflow cavities in the east-west direction extending $\sim$10,000~AU in radius \citep{Tobin2008}. On $\sim$2000~AU scales the protostellar envelope has a flattened shape with an infalling rotating velocity profile \citep{Ohashi1997}, and the presence of a rotationally supported disk with a $\sim$100~AU radius was derived from the continuum and $^{13}$CO emission \citep{Tobin2012}. Analysis of C$^{18}$O emission confirmed the presence of a Keplerian disk with a kinematic estimate of $\sim$74~AU for the disk radius \citep[][A17]{Ohashi2014}. The disk is seen almost edge-on ($i = 85^{\circ}$, \citealt{Tobin2008,Oya2015}), enabling direct probes of the midplane as shown for the Flying Saucer protoplanetary disk \citep{Dutrey2017}.

T13 modeled the (sub)millimeter continuum emission and visibilities, scattered light L$^{\prime}$ image, spectral energy distribution, and mid-infrared spectrum using 3D radiative transfer codes \citep{Whitney2003,Brinch2010}. Their best-fit model has a highly flared disk ($H \propto R^{1.3}$, with $H$ = 100 AU at a 48 AU radius) with a radius of 125~AU. The temperature structure has a midplane temperature of 30~K at 100~AU, implying that CO will not freeze out in the disk. Given the edge-on configuration, high resolution CO observations should be able to confirm or refute the presence of gas-phase CO throughout the entire disk midplane. These results can be reinforced by observations of molecular tracers of CO freeze-out like N$_2$D$^+$, which will be destroyed as long as CO is present in the gas phase. 

In this paper we use archival ALMA observations of CO isotopologues ($^{13}$CO and C$^{18}$O) and N$_2$D$^+$ to determine whether the embedded disk around L1527 resembles a Class II disk with a large cold gas reservoir, or whether this young disk is warm enough to have gaseous CO throughout. The data reduction is described in Sect.~\ref{sec:Observations}, and the resulting $^{13}$CO and C$^{18}$O images and non-detection of N$_2$D$^+$ are presented in Sect.~\ref{sec:Results}. In Sect.~\ref{sec:Analysis}, we use the optically thick $^{13}$CO and C$^{18}$O emission to derive a radial temperature profile for the disk midplane, and discuss the difficulties of disentangling disk and envelope emission. The effect of CO freeze out over different radial extents on the $^{13}$CO, C$^{18}$O and N$_2$D$^+$ emission is examined in Sect.~\ref{sec:Models} using the 3D radiative transfer code LIME \citep{Brinch2010}. In addition these models are used to address the CO abundance in the disk. Implications of the absence of CO freeze-out in embedded disks are discussed in Sect.~\ref{sec:Discussion}, and the conclusions are summarized in Sect.~\ref{sec:Conclusions}.  

\begin{figure*}
\centering
\includegraphics[width=\textwidth,trim={0cm 0.3cm 0cm 0.3cm},clip]{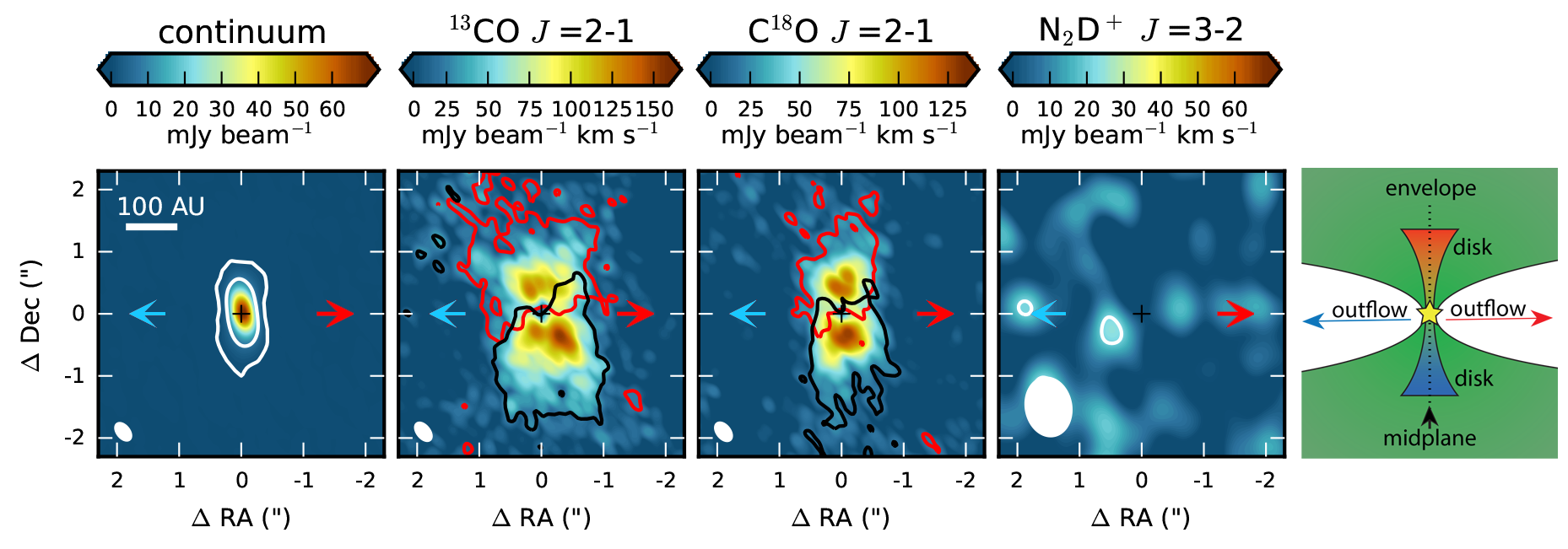}
\caption{Continuum image toward L1527 at 1.3 mm (\textit{first panel}) and integrated intensity maps for the $^{13}$CO $J=2-1$ (\textit{second panel}), C$^{18}$O $J=2-1$ (\textit{third panel}) and N$_2$D$^+$ $J=3-2$ (\textit{fourth panel}) transitions. For $^{13}$CO and C$^{18}$O, only channels with a $>$3$\sigma$ detection are included. The red and black contours indicate the 3$\sigma$ level of the red- and blueshifted emission, which corresponds to 18 and 21~mJy~beam$^{-1}$~km~s$^{-1}$ for $^{13}$CO, and 17 and 18~mJy~beam$^{-1}$~km~s$^{-1}$ for C$^{18}$O. The white contours denote the 5$\sigma$ and 50$\sigma$ levels (1$\sigma$ = 0.13 mJy beam$^{-1}$) for the continuum, and the 3$\sigma$ (20~mJy~beam$^{-1}$~km~s$^{-1}$) level for N$_2$D$^+$. The position of the continuum peak ($\alpha$(J2000)~=~04$^{\rm{h}}$39$^{\rm{m}}$53$\fs$88; $\delta$(J2000)~=~26$\degr$03$\arcmin$09$\farcs$57) is marked by a black cross and the blue and red arrows show the direction of the outflow. The beam is shown in the lower left corner of each panel. The \textit{right panel} shows a schematic view of the disk and inner envelope. The midplane is indicated by a dotted line.}
\label{fig:OverviewM0}
\end{figure*}



\section{Observations} \label{sec:Observations}

The L1527 data analyzed here were retrieved from the ALMA archive (project code: 2013.1.01086.S, PI: S. Koyamatsu). Observations were carried out during Cycle 2 on 24 May 2015, using 34 antennas sampling baselines between 21 and 539~m, and on 20 September 2015, using 35 antennas sampling baselines between 41 and 2269 m. This corresponds to a maximum angular scale of 4.2$\arcsec$ and 1.3$\arcsec$, respectively. The phase center was $\alpha$(J2000)~=~04$^{\rm{h}}$39$^{\rm{m}}$53$\fs$91; $\delta$(J2000)~=~26$\degr$03$\arcmin$09$\farcs$80. During both observations, L1527 was observed with a total on-source integration time of 25 minutes. The correlator setup included a 2 GHz continuum band centered at 233.0 GHz, and spectral windows with a bandwidth of 58.6 MHz targeting the $^{13}$CO $J=2-1$, C$^{18}$O $J=2-1$ and N$_2$D$^+$ $J=3-2$ transitions. The spectral resolution was 30.5~kHz for C$^{18}$O and 61.0~kHz for the two other lines, which corresponds to a velocity resolution of 0.04 and 0.08~km~s$^{-1}$, respectively. 

Calibration of the visibilities observed in the compact configuration was done using the ALMA Pipeline and version 4.2.2 of Common Astronomy Software Applications (CASA), while the observations in the more extended configuration were reduced manually by the East Asia ALMA Regional Center staff using CASA version 4.5. Calibrators for the respective observations were J0510+1800 and J0429+2724 (phase), J0423-0120 and J0510+1800 (bandpass and flux). We performed self-calibration on the continuum data and applied the phase and amplitude solutions also to the spectral line bands. The data were imaged using the CASA task \texttt{CLEAN} with natural weighting. For the line images, the continuum was subtracted before imaging. The total 1.3 mm continuum flux density derived from a Gaussian fit to the visibilities is 184.2 $\pm$ 0.15 mJy, consistent with the 1.3 mm flux density reported by A17. The natural weighted continuum image has a peak flux density of 66.0 $\pm$ 0.16 mJy beam$^{-1}$ (for a $0\farcs30 \times 0\farcs18$ beam), and is shown in Fig.~\ref{fig:OverviewM0}.  An overview of the image parameters for the observed molecular lines is provided in Table~\ref{tab:Lineparameters}. For $^{13}$CO and C$^{18}$O, only the observations in the extended configuration are used, while the N$_2$D$^+$ upper limits are derived from the data obtained in the more compact configuration.


\section{Results} \label{sec:Results}

Figure~\ref{fig:OverviewM0} shows the integrated intensity (zeroth moment) maps for $^{13}$CO ($J=2-1$), C$^{18}$O ($J=2-1$), and N$_2$D$^+$ ($J=3-2$) toward L1527, as well as a schematic overview of this region with the edge-on disk and inner part of the infalling rotating envelope. Carbon monoxide (CO) emission has been widely used to study the regions in protostellar systems where the material is warm enough ($\gtrsim$20~K) to evaporate CO off the dust grains. Emission lines from the main isotopologue $^{12}$C$^{16}$O are optically thick and dominated by the outflow material, while the less abundant isotopologues are relatively more optically thin and expected to probe the disk forming region. Meanwhile, the cold regions ($T \lesssim$~20~K) where CO is frozen out are traced by N$_2$H$^+$ and its deuterated variant N$_2$D$^+$ \citep[e.g.,][]{Jorgensen2004,Crapsi2005,Tobin2013b}.  

$^{13}$CO emission was detected above the 3$\sigma$ level at relative velocities $\Delta v$ with respect to the $v_{\rm{LSR}}$ of the source (5.9 km~s$^{-1}$), between -3.2 and -0.3 km~s$^{-1}$, and between 0.7 and 2.8 km~s$^{-1}$. C$^{18}$O emission was detected at similar velocities: between -3.0 and -0.3 km~s$^{-1}$, and between 0.5 and 2.8~km~s$^{-1}$. Line emission is not detected around the systemic velocity due to large-scale emission being resolved out. The zeroth moment maps presented in Fig.~\ref{fig:OverviewM0} are constructed using only channels with a $>3\sigma$ detection. Both CO isotopologues show emission elongated in the direction perpendicular to the outflow, with redshifted emission located north of the continuum peak and blueshifted emission to the south. This is consistent with a Keplerian rotating disk and the inner part of a infalling rotating envelope (\citealt{Tobin2012}, T13, \citealt{Ohashi2014}, A17). Estimates of the Keplerian disk radius vary between $\sim$74 and $\sim$125~AU (A17, T13, resp.), that is, $0\farcs5 - 0\farcs9$. The $^{13}$CO isotopologue is more abundant than C$^{18}$O which results in the $^{13}$CO emission being more extended in the east-west direction while C$^{18}$O is only detected in the denser regions toward the midplane. The line emission does not peak on source, probably because the continuum becomes optically thick in the innermost region around the protostar ($R \lesssim$15~AU, see also the discussion in Sect.~\ref{sec:Disc_model}), lowering the emission peak. 

The edge-on orientation of the disk allows for direct observation of the midplane, and thus, theoretically, for a direct determination whether CO is frozen out or present in the gas phase. From these observations, no obvious reduction in $^{13}$CO or C$^{18}$O emission, corresponding to CO freeze-out, is seen toward the midplane. Some asymmetry is visible, especially for $^{13}$CO, with the southern half of the disk being brightest west of the midplane and the northern half east of the midplane. This asymmetry is, at least partly, the result of the $uv$-coverage of the observations. Similar asymmetrical features can be introduced for axisymmetric models simulated with the observed visibilities. 

In contrast with the clear CO isotopologue detections, no emission was detected for N$_2$D$^+$, which can only be abundant when CO is frozen out (at $T \lesssim$ 20~K and $n \gtrsim 10^{5}$~cm$^{-3}$). The 3$\sigma$ limit of 20~mJy~beam~km~s$^{-1}$ corresponds to an upper level for the beam-averaged N$_2$D$^+$ column density of $\sim 2-4 \times 10^{12}$~cm$^{-2}$ (assuming LTE and excitation temperatures between 10 and 50~K; see e.g., \citealt{Goldsmith1999}). Thus, at first inspection there is no evidence in these observations for a large cold outer region in the disk around L1527 where CO is frozen out.


\section{Analysis} \label{sec:Analysis}

To examine the temperature structure of the L1527 disk in more detail we analyse the $^{13}$CO and C$^{18}$O emission. First, we establish that emission from both molecules is optically thick using the line ratio (Sect.~\ref{sec:OptDepth}). Next, we derive a temperature profile for the disk midplane (Sect.~\ref{sec:Temp}). Although the edge-on configuration of the disk allows a direct view of the midplane, deriving a radial temperature profile is not straightforward. The $^{13}$CO and C$^{18}$O emission trace both the disk and inner envelope, so emission originating from the disk has to be disentangled from envelope emission. In addition, the radius from which the optically thick emission originates has to be determined. Viewing the disk edge-on means that the line of sight goes through the entire midplane before reaching the inner most part. If the emission becomes optically thick already in the inner envelope or outer disk, the inner disk cannot be observed. Moreover, this means that on-source emission does not necessarily originate at small radii ($\lesssim 40$ AU). Before addressing these problems in detail using the velocity structure (Sect.~\ref{sec:DiskEnvelope}), we derive a radial temperature profile in a conservative way. That is, we measure the temperature in the disk midplane at the highest possible velocity offsets. Assuming a power law temperature profile we can then estimate the midplane temperature structure for the entire disk (Sect.~\ref{sec:Temp}).

\begin{figure}
\centering
\includegraphics[width=\textwidth,trim={0cm 15cm 0cm 0.8cm},clip]{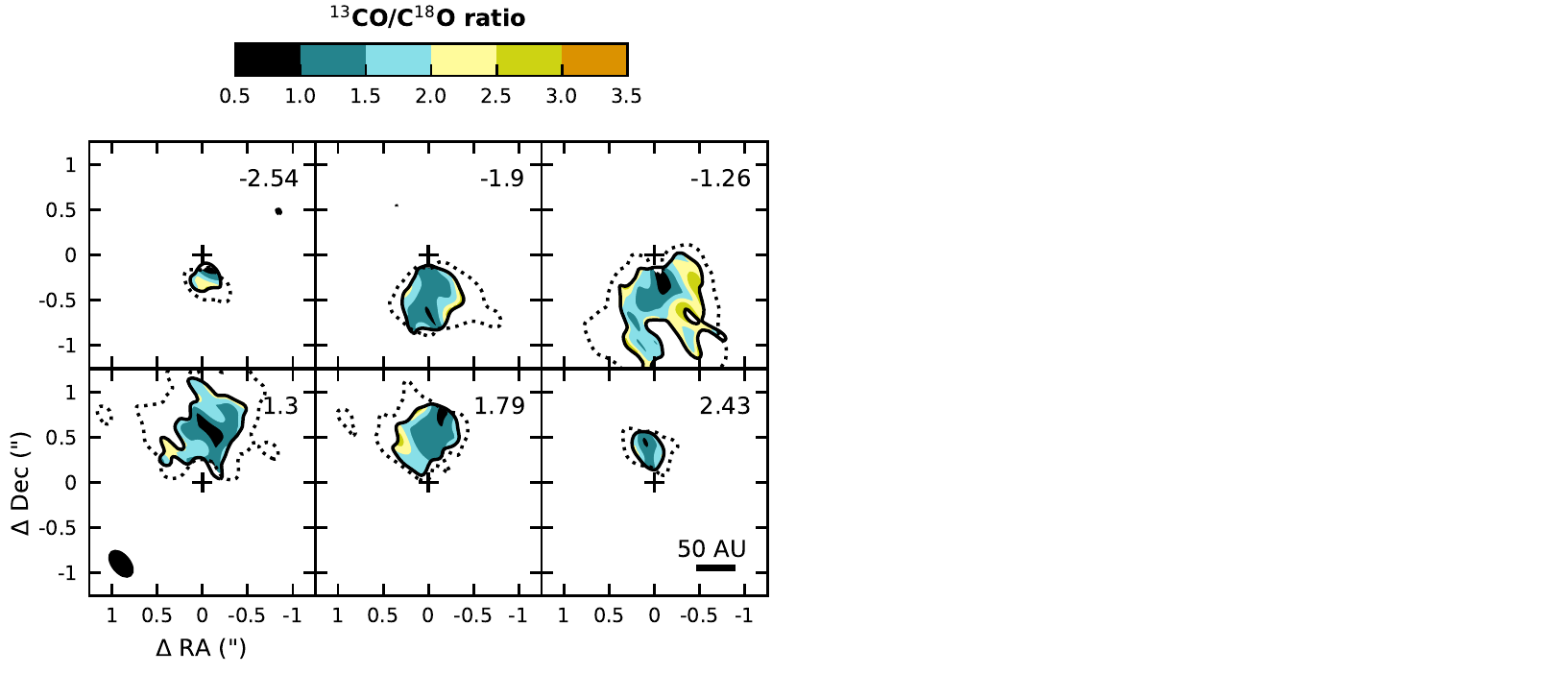}
\caption{$^{13}$CO ($J=2-1$)/C$^{18}$O ($J=2-1$) line intensity ratio in six representative velocity channels. The velocity relative to the systemic velocity of $v_{\rm{lsr}} = 5.9$ km s$^{-1}$ is listed in the top right corner of each panel and the channel width is 0.16~km~s$^{-1}$. The line ratio is only calculated in pixels with at least a $3\sigma$ detection for both $^{13}$CO and C$^{18}$O. The dotted and solid contours mark the respective $3\sigma$ levels. The plotted spatial scale is such as to highlight the inner $\sim 1\arcsec$ (140 AU), that is, the disk and inner envelope. The continuum peak position is indicated with a cross and the synthesized beam is shown in the lower left corner of the lower left panel.}
\label{fig:Lineratio}
\end{figure}

\subsection{$^{\sf13}$CO optical depth} \label{sec:OptDepth}

Given the low critical density of the $^{13}$CO $J=2-1$ transition ($n_{\rm{crit}} \sim 10^4$~cm$^{-3}$), the emission is expected to be thermalized throughout the disk and in the inner envelope ($n > 10^6$~cm$^{-3}$). This means that if the emission is optically thick, the brightness temperature is a measure of the excitation temperature, which in turn is a lower limit of the kinetic temperature of the gas. The optical depth of the $^{13}$CO emission can be derived from the $^{13}$CO/C$^{18}$O line ratio. If both lines are optically thin, the line ratio will be equal to the abundance ratio of $^{13}$CO and C$^{18}$O ($\sim$7--8), given no isotope-selective photodissociation processes. This is a valid assumption for the disk midplane, since UV radiation will only penetrate into the disk surface layers. As the $^{13}$CO emission becomes optically thick, the line ratio decreases. The line ratio becomes unity when both lines are optically thick.  

The $^{13}$CO ($J=2-1$) / C$^{18}$O ($J=2-1$) line ratio is calculated for each pixel with at least a $3\sigma$ detection (see Table~\ref{tab:Lineparameters}) for both $^{13}$CO and C$^{18}$O, and six representative channels are presented in Fig.~\ref{fig:Lineratio}. The $^{13}$CO/C$^{18}$O ratio is below 3.0 in all regions, and below 2.0 in the pixels with the highest signal-to-noise (along the midplane). The $3\sigma$ uncertainty on the $^{13}$CO/C$^{18}$O ratio is calculated for each pixel by propagating the rms in the 0.16~km~s$^{-1}$ channels. Within this uncertainty, the ratio remains below 3.5, except in a few pixels close to the $3\sigma$ contour of the C$^{18}$O emission. Theoretically, the ratio cannot be smaller than 1.0, unless $^{13}$CO emission is significantly more resolved out relative to C$^{18}$O. The lowest values reached are $\sim$0.5, but the $3\sigma$ uncertainty in these pixels is $\geq$0.5. The values below 1.0 are thus not significant.

\begin{figure*}
\centering
\includegraphics[width=\textwidth,trim={0cm 15.4cm 0cm 1.0cm},clip]{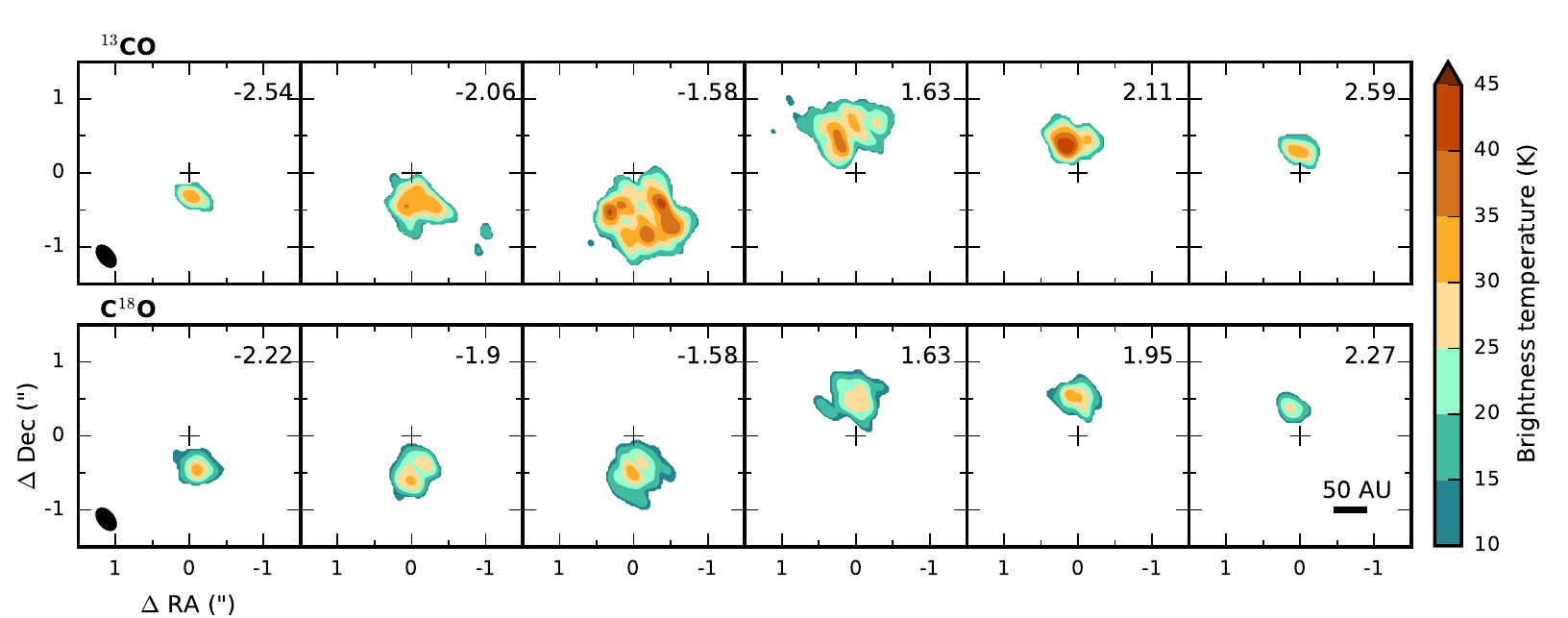}
\caption{Brightness temperature for the $^{13}$CO $J=2-1$ (\textit{top panels}) and C$^{18}$O $J=2-1$ (\textit{bottom panels}) transitions in six representative velocity channels. Channel velocities with respect to the systemic velocity of $v_{\rm{lsr}} = 5.9$ km s$^{-1}$ are listed in the top right corner of each panel and the channel width is 0.16~km~s$^{-1}$. All channels are shown in Figs.~\ref{fig:13CO_TB_all} and \ref{fig:C18O_TB_all}. The continuum peak position is marked with a cross and the beam is shown in the lower left corner of the left panels.}
\label{fig:TB}
\end{figure*}


Since the optical depth depends on the gas temperature, the quantity we set out to determine, the line ratio is also dependent on the temperature. Therefore, in order to determine whether the emission from either or both CO isotopologues is optically thick, we calculate line ratios for a range of temperatures (20--100 K) and column densities (C$^{18}$O column densities ranging from $10^{13}$~cm$^{-2}$ to $10^{18}$~cm$^{-2}$) using RADEX\footnote{\texttt{http://home.strw.leidenuniv.nl/}$\sim$\texttt{moldata/radex.html}} \citep{vanderTak2007,vanderTak2011} with collisional rate coefficients from \citet{Yang2010}. The isotope ratios are taken to be [$^{12}$C]/[$^{13}$C] = 77 and [$^{16}$O]/[$^{18}$O] = 560 \citep{Wilson1994}, which gives an abundance ratio of $\sim$ 7 for $^{13}$CO and C$^{18}$O. For all temperatures considered, a $^{13}$CO/C$^{18}$O ratio below 4 can only be reached if the $^{13}$CO emission is optically thick (Fig.~\ref{fig:RADEXLineratio}, left panel). When the C$^{18}$O emission becomes optically thick as well, the line ratio drops below 1.5 (Fig.~\ref{fig:RADEXLineratio}, right panel). Taking [$^{12}$C]/[$^{13}$C] = 70, that is, [$^{13}$CO]/[C$^{18}$O] = 8, does not change these results. This thus means that the observed $^{13}$CO emission is generally optically thick ($\tau > 1$), except maybe in the upper most layers of the disk. In addition, C$^{18}$O becomes optically thick in the densest regions near the midplane where $^{13}$CO/C$^{18}$O < 1.5.

\subsection{Gas temperature} \label{sec:Temp}

\begin{figure}
\centering
\includegraphics[trim={0cm 6cm 0cm 1.0cm},clip]{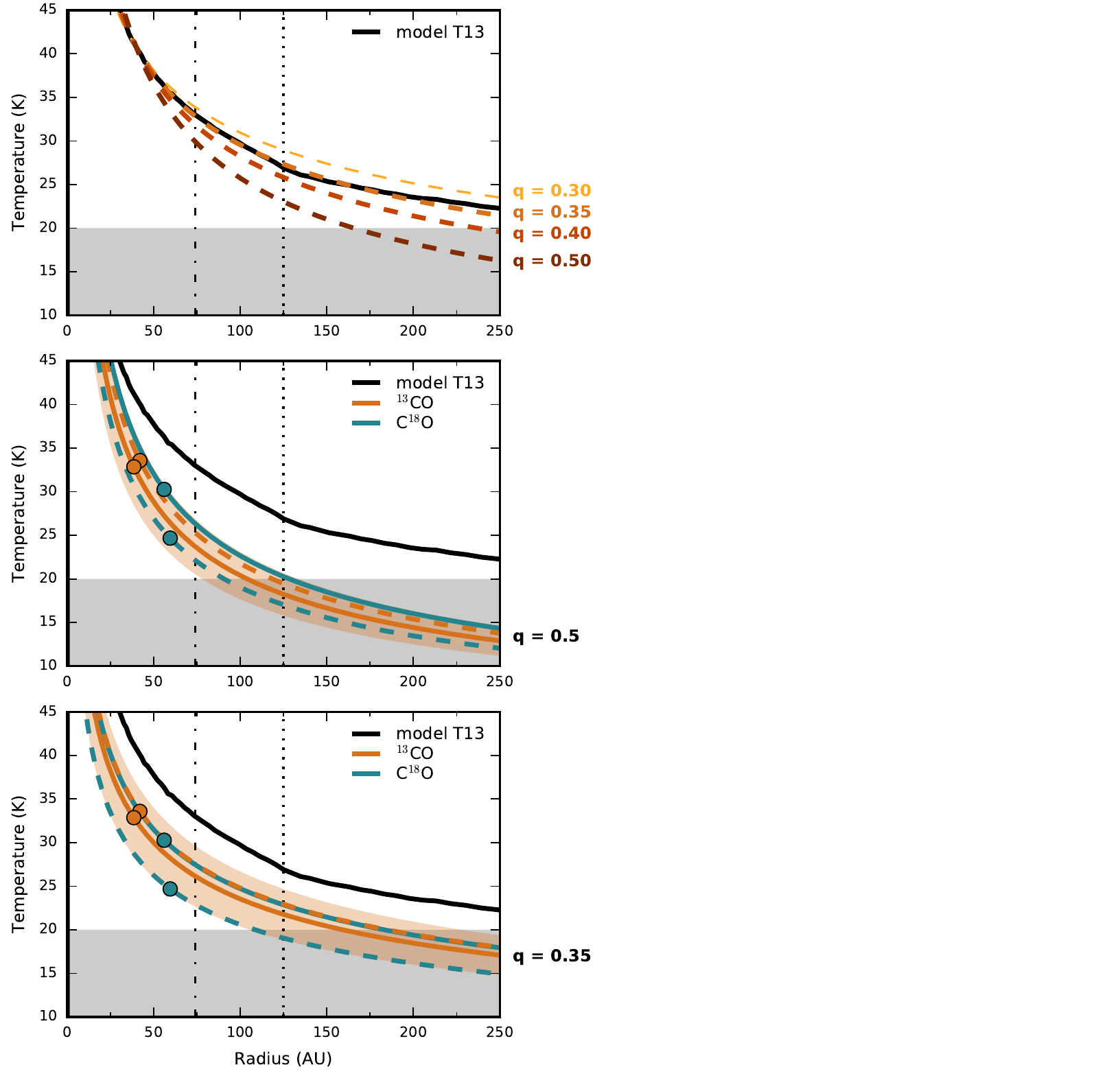}
\caption{Radial midplane temperature profiles for the L1527 disk and inner envelope. The \textit{top panel} shows a comparison between the temperature profile from the T13 model and power laws with different exponents ($q$). In the \textit{middle} and \textit{bottom panels}, power law profiles are derived from $^{13}$CO (orange lines) and C$^{18}$O observations (blue lines) assuming $q=0.5$ and $q=0.35$, respectively. Solid lines are derived from a redshifted channel, dashed lines from a blueshifted channel. The measured values are indicated by filled circles. The orange shaded area shows the 1$\sigma$ uncertainty on the solid orange line ($^{13}$CO from redshifted channel). The gray shaded area marks the temperature range for which CO is expected to be frozen out (i.e., below 20 K). The vertical dash-dotted line marks a disk radius of 74~AU as determined by A17, and the dotted line a disk radius of 125~AU derived by T13.}
\label{fig:T_profile}
\end{figure}


Since the $^{13}$CO and C$^{18}$O emission is optically thick ($\tau > 1$), their brightness temperatures can be used as a measure for the kinetic temperature of the gas. The brightness temperature for six representative channels is presented in Fig.~\ref{fig:TB}. All channels are displayed in Appendix~\ref{ap:Tb}. For $^{13}$CO, the temperature along the midplane is above $\sim$25~K, except at $|\Delta v| \geq 2.7$~km~s$^{-1}$, where the emission is unresolved, and around $|\Delta v| \sim 1$~km~s$^{-1}$, where there may still be an effect of emission being resolved out. The C$^{18}$O emission is weaker than the $^{13}$CO emission, but the channels with resolved emission show the same result: the midplane temperature is above $\sim$25~K. 

Despite the complexity of locating the radial origin of the emission (see Sect.~\ref{sec:DiskEnvelope}), we can make a conservative estimate of the radial temperature profile by determining the maximum temperature in the highest velocity channels and extrapolating this assuming a radial power law dependence,
\begin{equation}
T = T_{\rm{obs}}(R/R_{\rm{obs}})^{-q},
\end{equation}
where $T_{\rm{obs}}$ and $R_{\rm{obs}}$ are the temperature and radius derived from the observations. Since the highest velocities are reached closest to the star, these channels are the least likely to be contaminated by envelope emission.  Often a power law exponent of $q=0.5$ is used, appropriate for a black body flared disk with scale height $h(r) = h_0(r/R_\star)^{1.25}$ \citep{Kenyon1987}, but $q$ can range between 0.33 and 0.75 \citep[see e.g.,][]{Adams1986,Kenyon1993,Chiang1997}. However, full radiative transfer modeling shows that the temperature can not be characterized by a single power law, and that the midplane temperature profile can be flatter when the disk is embedded in an envelope \citep[e.g.,][]{Whitney2003}. We therefore compared a power law with $q=0.5$ to the midplane temperature profile from the T13 best fit model (see Fig.~\ref{fig:T_profile}, top panel). T13 modeled the disk continuum emission by fitting both the visibilities and images of 870 $\mu$m and 3.4 mm observations, the multi-wavelength SED and $L^{\prime}$ scattered light images with 3D radiative transfer modeling (see Appendix~\ref{ap:Model} for a more detailed description). A power law with $q=0.5$ appears to be steeper than the T13 profile, while adopting $q=0.35$ results in a better match to the T13 model between 40 and 125~AU. Assuming $q=0.5$ will thus likely provide a lower limit to the temperature in the outer disk.

For $^{13}$CO, the largest observed velocity offsets containing resolved emission are $\Delta v$ = -2.54 and $\Delta v$ = 2.59 km~s$^{-1}$. The blueshifted channel has a maximum temperature of 34~K along the midplane at 42~AU (0.3$\arcsec$ from the continuum peak position), and in the redshifted channel this is 33~K at 39~AU. Even with the rms of 4~K~channel$^{-1}$, the temperature is thus above 20~K around 40~AU within the 3$\sigma$ uncertainty. The onset of CO freeze-out is therefore expected at radii $>$40~AU. Figure~\ref{fig:T_profile} shows the resulting radial profiles assuming $q=0.5$ (middle panel) and $q=0.35$ (bottom panel). For $q=0.5$, the temperature drops below 20~K between approximately 100 and 125~AU. Adopting the 125~AU disk radius derived by T13 this would mean that there could be a small cold outer region in the disk where CO is frozen out. A disk radius of 74~AU as determined by A17 would however mean that the entire disk is too warm for CO freeze-out. It is also expected that CO is not frozen out in a 125~AU disk for $q=0.35$.

For C$^{18}$O, the highest velocity offsets that show resolved emission are $\Delta v$~=~-2.22 and $\Delta v$~=~2.43~km~s$^{-1}$. The constructed temperature profiles are very similar to those derived from $^{13}$CO (Fig.~\ref{fig:T_profile}). The temperature in the blueshifted channel is probably lower because the emission does not peak along the midplane (see Fig.~\ref{fig:TB}). Since the C$^{18}$O emission is more confined to the midplane than $^{13}$CO, which extends to larger heights above the midplane, the similarity in the temperature profiles suggests that the temperature measured in the midplane is not strongly affected by blending with the warmer upper layers. The observed temperatures are lower than expected based on the T13 model. In Sect.~\ref{sec:Models_CO}, we will discuss that the measured brightness temperature, and consequently the corresponding radial temperature profile, underestimates the kinetic gas temperature. In summary, extrapolating the temperature measured for the highest velocity material in the inner disk ($\lesssim$40~AU) suggest that the temperature is likely to remain $\gtrsim$20~K out to at least $\sim$75~AU, and possibly out to $\sim$150~AU. 

The CO freeze-out temperature is $\sim$20~K when CO binds to pure CO ice. When CO binds to a water ice surface the freeze-out temperature is higher ($\sim$25~K, see e.g., \citealt{Burke2010}, \citealt{Fayolle2016}). Assuming $q=0.5$, this would mean that the snowline is at radii $\gtrsim$ 50~AU, allowing CO freeze-out in the outer disk also for a radius of 74~AU. With a power law dependence similar to the T13 profile ($q=0.35$), the disk remains $\gtrsim$25~K out to $\sim$85~AU. Thus, with measurements of the disk radius varying between 74 and 125~AU, these results indicate that a large fraction of the disk, or even the entire disk, is warm enough to prevent CO freeze-out.

\subsection{Disentangling disk and envelope emission} \label{sec:DiskEnvelope}

As mentioned above, determining the radial origin of the optically thick CO isotopologue emission is not trivial as both the disk and inner envelope contain CO. In addition, because the emission originates at the radius at which it becomes optically thick, the observed angular offset does not necessarily correspond to the physical radial offset. These problems, and their effect on the inferred power law temperature profile, can be investigated using the velocity structure of the system. To do so, we calculate the velocity along the line of sight for each position in the midplane. Then, for each 0.16~km~s$^{-1}$ velocity bin of the observations, we select the material moving at that velocity range and calculate the molecular column density along the line of sight. Finally, we convert this column density $N$ into optical depth $\tau$ and locate the $\tau = 1$ surface: 
\begin{equation}
\tau = \frac{c^3}{8\pi\nu^3}\frac{A_{\mathrm{ul}}}{\Delta V}\frac{g_{\mathrm{up}}N}{Q} \exp \left( \frac{-E_{\mathrm{up}}}{kT} \right) \left( \exp \left( \frac{h\nu}{kT} \right) - 1\right), 
\end{equation} 
where $\nu$ and $A_{\rm{ul}}$ are the frequency and the Einstein-A coefficient of the transition, respectively, $g_{\rm{up}}$ and $E_{\rm{up}}$ are the degeneracy and energy of the upper level, respectively, $Q$ is the partition function and $\Delta V$ is the line width (FWHM).  

We adopt the physical structure (dust density and temperature) from T13, who modeled the disk continuum emission by fitting both the visibilities and images of 870 $\mu$m and 3.4 mm observations, the multi-wavelength SED and $L^{\prime}$ scattered light images with 3D radiative transfer modeling. Their best fit model has a disk radius of 125~AU. Inside this radius the velocity is Keplerian, while the material in the envelope has a infalling rotating velocity profile \citep{Ulrich1976,Cassen1981}. More details can be found in Appendix \ref{ap:Model}. A critical parameter for the velocity structure is the stellar mass. \citet{Tobin2012} derived a dynamical mass of $\sim$0.19 $M_{\sun}$, while recently A17 find a slightly larger value of $\sim$0.45 $M_{\sun}$. Here we adopt the higher value so our estimate of the channels containing only disk emission is conservative; for lower stellar masses, envelope emission will be confined to smaller velocity offsets. We will refer to this as our fiducial model. To calculate the optical depth we assume a gas-to-dust ratio of 100 and a constant canonical CO/H abundance of $10^{-4}$ in the region with temperatures larger than 20~K. We assume the gas temperature is equal to the dust temperature, which is a valid approximation in the dense midplane.   

\begin{figure}
\centering
\includegraphics[width=\textwidth,trim={0.3cm 16cm 0cm 0.1cm},clip]{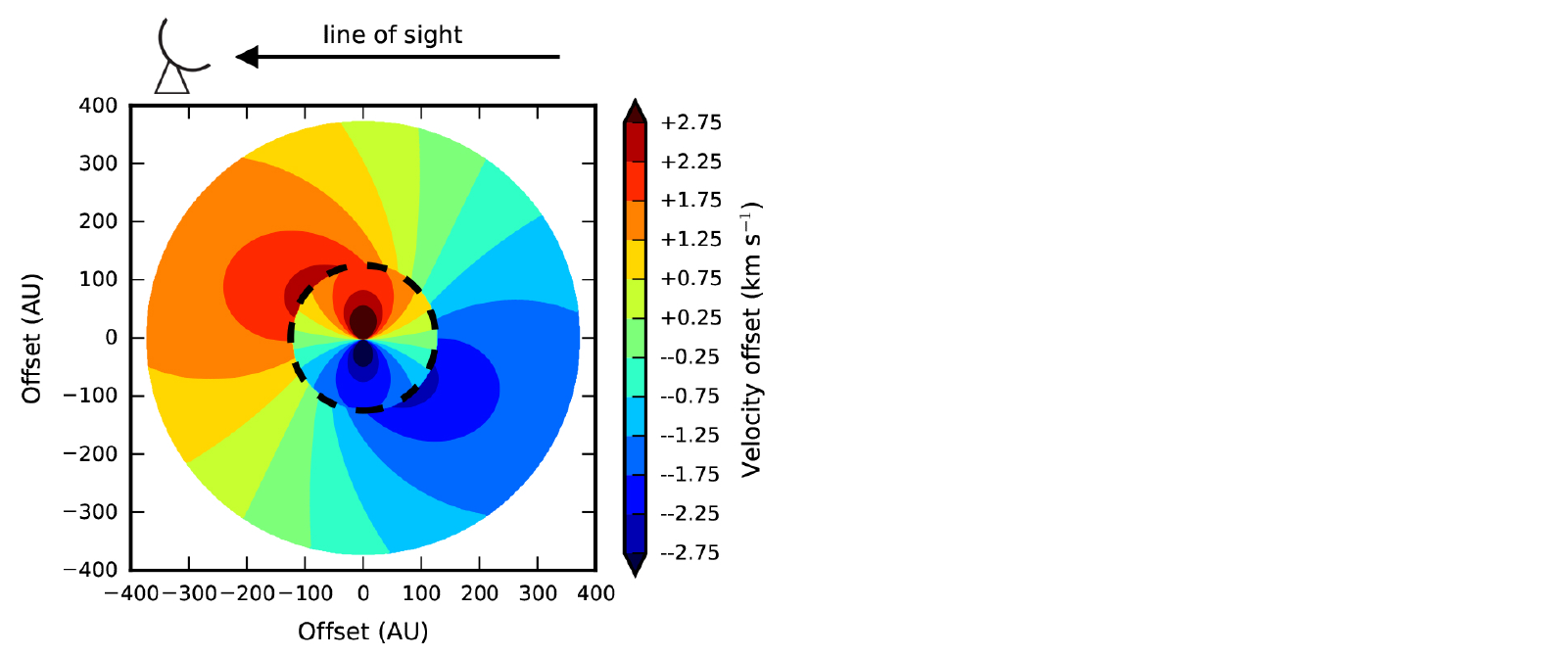}
\caption{Face-on view of the midplane in the disk and inner envelope for the fiducial model, showing the velocity component along the line of sight (indicated by the black arrow on top of the panel; the observer is on the left). The envelope material has a infalling rotating velocity profile \citep{Ulrich1976,Cassen1981}, while the disk material is rotating with Keplerian velocity. The adopted disk outer radius of 125~AU is marked by the dashed line. The outer edge of the envelope is set at the radius where the temperature in the best-fit model from T13 drops below 20~K, i.e. the CO freeze-out temperature. The adopted stellar mass is 0.45 $M_{\sun}$ (A17).}
\label{fig:Midplane}
\end{figure}

The resulting midplane velocity profile projected along the line of sight for the fiducial model is shown in Fig.~\ref{fig:Midplane}, and the midplane temperature and density structure from T13 can be found in Fig.~\ref{fig:Midplane_phys}. For the above described parameters, velocities larger than $\sim \pm 2.5$~km~s$^{-1}$ can only be reached in the disk, so these velocity channels are expected to contain only disk emission (see Table~\ref{tab:Velocities}). For a 0.19~$M_{\sun}$ star, disk-only emission would be observed for velocity offsets larger than $\sim$1.7~km~s$^{-1}$. Assuming free-fall velocity for the envelope material gives similar results (see Fig.~\ref{fig:Midplane_ff}). A17 use the modified free fall velocity structure from \citet{Ohashi2014}, that is, the free fall velocity reduced by a factor of 0.3, and derive a disk radius of 74~AU. For these parameters, channels with velocity offsets larger than $\sim$1.1~km~s$^{-1}$ would contain emission originating only in the disk.

\begin{table}
\addtolength{\tabcolsep}{-2pt}
\caption{Origin of emission at different velocity offsets.
\label{tab:Velocities}} 
\centering
\begin{tabular}{l c c c}
    \hline\hline
    \\[-.3cm]
    
    Envelope   & $M_{\rm{star}}$ & Disk only     & Envelope only\tablefootmark{a} \\ 
    velocity   & ($M_{\sun}$)      & (km s$^{-1}$) & (km s$^{-1}$) \\
    \hline 
    \\[-.3cm]
    UCM\tablefootmark{b} & 0.19 & $v \leq -1.74; v \geq 1.79$ & $0.98 \leq v \leq 1.30$ \\
    \textbf{UCM\tablefootmark{b}} & \textbf{0.45} & $\mathbf{v \leq -2.54; v \geq 2.59}$ & $\mathbf{1.47 \leq v \leq 1.79}$ \\
    Free fall            & 0.19 & $v \leq -1.74; v \geq 1.79$ & $v = 0.98$ \\
    Free fall            & 0.45 & $v \leq -2.54; v \geq 2.59$ & $1.47 \leq v \leq 1.63$ \\
    Modified \\
    \hspace{0.1cm} free fall\tablefootmark{c} & 0.45 & $v \leq -1.10; v \geq 1.14$ & $0.50 \leq v \leq 0.98$ \\
    \hline
\end{tabular}
\tablefoot{The fiducial model is highlighted in boldface. A disk radius of 125 AU is adopted for all models, except in the modified free fall case where a radius of 74 AU is used (A17). For a 125 AU disk, the disk-only channels would then be $v \leq -0.94; v \geq 0.98$ km s$^{-1}$ and the envelope-only channels would be $0.50 \leq v \leq 0.66$ km s$^{-1}$.  \tablefoottext{a}{Only for $^{13}$CO for which the redshifted emission becomes optically thick in the envelope.} \tablefoottext{b}{Infalling rotating envelope \citep{Ulrich1976,Cassen1981}.} \tablefoottext{c}{Free fall velocity modified by a factor 0.3 and a disk radius of 74 AU \citep[][A17]{Ohashi2014}.}}
\end{table}

\begin{figure*}
\centering
\includegraphics[width=\textwidth]{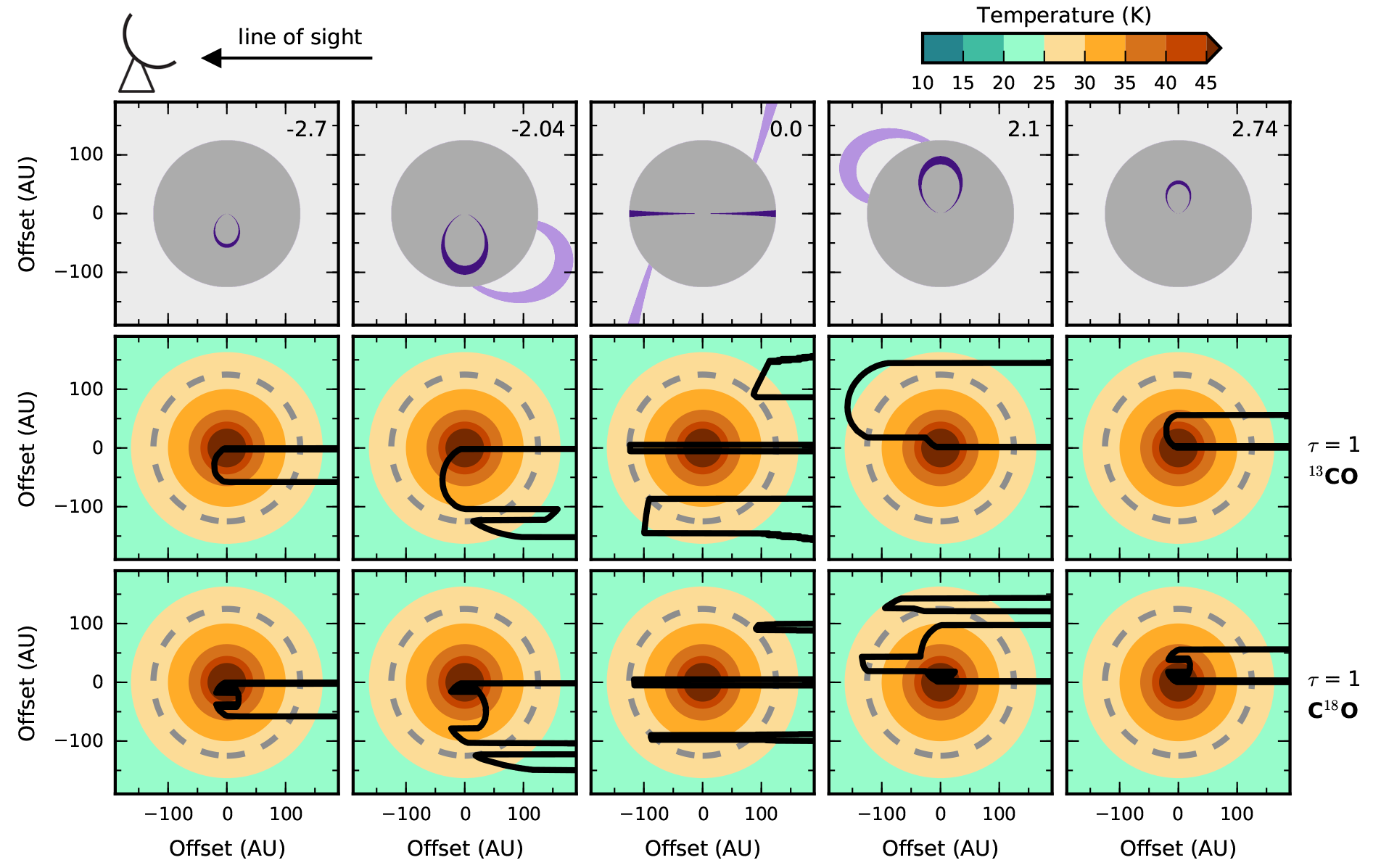}
\caption{Five representative velocity channels to illustrate where in the system the $^{13}$CO and C$^{18}$O emission originates for the fiducial model with a constant CO/H abundance of $10^{-4}$. The velocity profile in the disk is Keplerian and the envelope material is rotating and infalling (see Fig.~\ref{fig:Midplane}). The \textit{top panels} show the material moving at the line-of-sight velocity (with respect to the systemic velocity of 5.9~km~s$^{-1}$) indicated in the top right corner. Material in the disk (dark gray circle) is depicted in dark purple, envelope material in light purple. The solid black line in the \textit{middle and bottom panels} marks the $\tau = 1$ surface for an observer located on the left (as depicted by the arrow on top of the figure) for $^{13}$CO and C$^{18}$O, respectively. The color scale represents the temperature in the best-fit model from T13 and the dashed line indicates the adopted disk outer radius of 125 AU.}
\label{fig:Optdepth}
\end{figure*}

For a 0.45 $M_{\sun}$ star, both the disk and envelope contain material that has line-of-sight velocity offsets smaller than $\pm 2.5$~km~s$^{-1}$. What material is responsible for the emission thus depends on where the emission becomes optically thick. The $\tau = 1$ surface for $^{13}$CO in five representative channels ($\Delta v =$ \mbox{-2.70}, -2.04, 0, 2.10 and 2.74~km~s$^{-1}$) is shown in Fig.~\ref{fig:Optdepth} (see Fig.~\ref{fig:Optdepth_ff} for the case of a free falling envelope). At blueshifted velocities the southern half of the disk is observed, as well as that part of the envelope that is behind the disk for our line of sight (see Fig.~\ref{fig:Midplane}). The emission becomes optically thick in the disk, so the envelope is blocked from our view except at velocities where the envelope extends further south than the disk. In this case, at small angular offsets from the source center the emission originates in the disk, while emission at larger offsets traces the envelope (see Fig.~\ref{fig:Optdepth}, second column). In the redshifted channels the emission becomes optically thick in the envelope between us and the disk. At the highest  velocities occuring in the envelope, disk emission will be present very close to the protostar (see Fig.~\ref{fig:Optdepth}, fourth column), while envelope emission is present at larger angular offsets. However, the angular offset of the envelope emission ($\sim$30~AU in this case) is smaller than the radius at which the emission originate ($\sim$125~AU). At lower velocities only envelope emission will be observed (see Table~\ref{tab:Velocities}), until at velocities close to the systemic velocity the disk can be seen at the source position with envelope emission on both sides (see Fig.~\ref{fig:Optdepth}, third column). The situation is similar for C$^{18}$O (see Fig.~\ref{fig:Optdepth}, bottom row), except that the emission becomes optically thick in the far side of the disk (see Fig.~\ref{fig:Optdepth}, column one, two and five), instead of the near side. In addition, the emission remains optically thin in most of the envelope at redshifted velocities (for the adopted density structure; see Fig.~\ref{fig:Optdepth}, fourth column). The disk is thus not hidden from our view.

For clarity, we have neglected the dust opacity in Fig.~\ref{fig:Optdepth}. Adopting the dust opacity from T13, that is, \mbox{$\kappa = 3.1$ cm$^2$ g$^{-1}$} at 220~GHz, the dust would only become optically thick in the inner $\sim$10~AU. If the velocity profile in the envelope is free fall (Fig.~\ref{fig:Midplane_ff}), the highest envelope velocities are found along the line of sight toward the protostellar position, instead of offset to the north (for redshifted emission) or to the south (for blueshifted emission), as in the infalling rotating case. This changes the location of the $\tau = 1$ contour at velocities that occur in both the disk and envelope (that is, columns two and four in Fig.~\ref{fig:Optdepth} will look different; see Fig.~\ref{fig:Optdepth_ff} and Table~\ref{tab:Velocities}). For blueshifted velocities, the disk will be visible south of the protostar, while envelope emission will be present to the north. For redshifted velocities, emission surrounding the protostar originates from within the envelope, while the emission furthest north is coming from the disk. 

In summary, for $M_{\star} =$ 0.45 $M_{\sun}$, channels with velocity offsets larger than $\sim$2.5 km s$^{-1}$ are expected to contain only the disk's emission. In contrast, channels with smaller velocity offsets show emission from both the disk and the inner envelope. The most important consequence of this is that the angular offset of optically thick emission with respect to the protostellar position in these channels does not always directly correspond to the radial position of the emitting material.

Determining exactly where the emission originates is non-trivial and requires knowledge of, amongst others, the disk radius, density structure and CO abundance. Nonetheless, a global view of the temperature in the disk and inner envelope can be obtained through the brightness temperature in the different velocity channels. As shown in Sect.~\ref{sec:Temp}, for both $^{13}$CO and C$^{18}$O the temperature along the midplane is above $\sim$25~K at all velocities, suggesting a warm disk and inner envelope. The highest velocity channels that contain resolved $^{13}$CO emission, and were used to construct the radial temperature profile, are $\Delta v$ = -2.54 and $\Delta v$ = 2.59 km~s$^{-1}$. For stellar masses $\leq 0.45$ $M_{\sun}$, these channels are indeed expected to trace only the disk. The velocity offsets used for C$^{18}$O are slightly smaller ($\Delta v$ = -2.22 and $\Delta v$ = 2.43 km~s$^{-1}$), and could be contaminated by envelope emission if the stellar mass is 0.45~$M_{\sun}$. However, unlike for $^{13}$CO, the redshifted C$^{18}$O emission is expected to become optically thick mostly in the disk and not already in the envelope. Moreover, the measured temperatures are similar as for $^{13}$CO.

Depending on the stellar mass and the envelope velocity structure, a number of redshifted velocity channels are predicted to contain only $^{13}$CO emission from the inner envelope (see Table~\ref{tab:Velocities}). For $M_{\star} \leq 0.45$ $M_{\sun}$, these channels have velocity offsets smaller than 1.79~km~s$^{-1}$. In all these channels the $^{13}$CO brightness temperature remains above $\sim$25~K (see Fig~\ref{fig:13CO_TB_all}). So if this emission indeed originates in the inner envelope, the temperature in that region would be higher than 25~K. However, the power law temperature profile predicts inner-envelope temperatures of $\sim$15--22~K. This could mean that there is a jump in the temperature between the disk and inner envelope. Such a jump, albeit stronger, has been suggested by \citet{Sakai2014a} based on observations of SO. Or alternatively, the temperature profile is less steep than shown in Fig.~\ref{fig:T_profile}, assuming the temperature derived for the inner disk is not underestimated. The inner disk temperature is determined from channels expected to contain only disk emission, and is thus not likely to be underestimated due to envelope contamination (see Fig.~\ref{fig:Optdepth}). A shallower temperature profile would mean a warmer disk with temperatures $\gtrsim$25~K for all radii (125 AU).


\section{Modeling of the line emission} \label{sec:Models}

\subsection{$^{13}$CO and C$^{18}$O} \label{sec:Models_CO}

\begin{figure}
\centering
\includegraphics[width=0.5\textwidth,trim={0.1cm 2.2cm 0cm 2.0cm},clip]{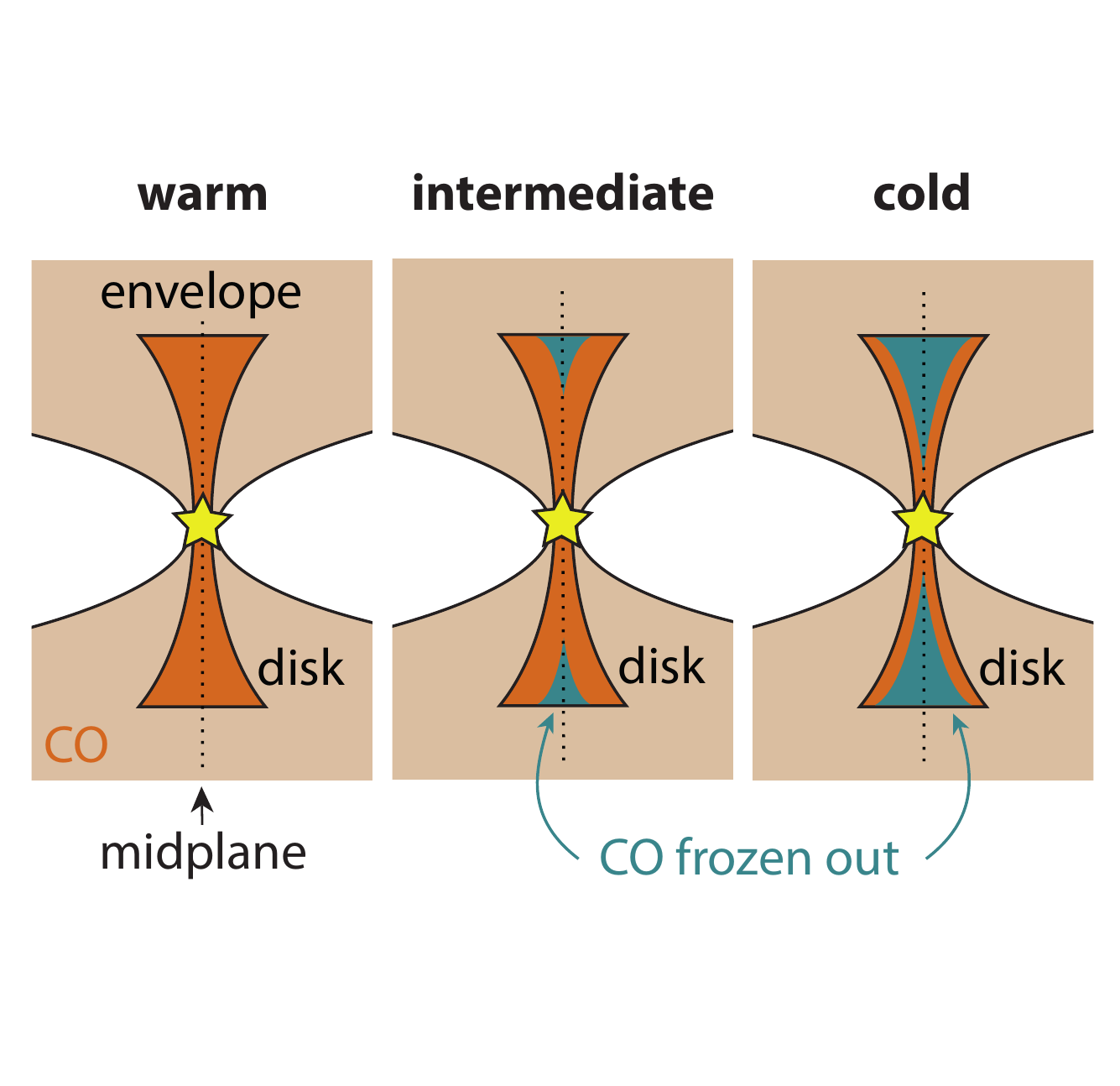}
\caption{Three different models for the CO distribution in the L1527 disk. \textit{Left panel}: a warm disk with no CO freeze out. \textit{Middle panel:} a slightly colder disk where CO is frozen out in the outer most region ($R > $ 71~AU) near the midplane. \textit{Right panel:} a cold disk where gaseous CO is only present in the inner most region ($R <$ 23~AU) and in the surface layers. CO is present in the gas phase in the inner envelope in all models.} 
\label{fig:Models}
\end{figure}

Based on the brightness temperature in the $^{13}$CO and C$^{18}$O channels, the disk and inner envelope are likely to have midplane temperatures above the CO freeze-out temperature of $\sim$20~K. However, a power law temperature profile based on the temperature at the highest velocity offsets suggests that there may still be a cold outer region in the disk. To study in more detail the effect of CO freeze-out on the emission, we simulate $^{13}$CO and C$^{18}$O emission for three different scenarios (see Fig.~\ref{fig:Models}). The focus lies on determining whether the observations can show the difference between a completely warm disk and a disk with a cold outer midplane. The first model (`warm model') consists of a warm disk that has no CO freeze-out. In the second case (`intermediate' model), there is a small region in the outer disk where CO is frozen out, and in the third model (`cold' model) the temperature is only high enough to have gaseous CO in the innermost region of the disk and in the surface layers. In all models CO is present in the gas phase in the inner envelope. Models without CO in the envelope cannot reproduce the angular extent of the observed emission. 

The exact distribution of CO in the different models is based on a simple parametrization: for $T \geq 20$~K CO is present in the gas phase at a constant abundance of $10^{-4}$ (w.r.t. H), while for $T < 20$~K CO is frozen out and the abundance is set to zero. For the warm model we adopt the temperature structure of the T13 model, because the temperature is above  $\sim$25~K in the entire disk (see Fig.~\ref{fig:2Dstructure}). For the intermediate and cold models, the T13 temperature is scaled down by a factor to create regions with $T < 20$~K (schematically shown in Fig.~\ref{fig:Models}). For the intermediate model the temperature is lowered by 40\% and for the cold model by 60\%, resulting in a CO snowline at 71~AU and 23~AU, respectively (see Fig.~\ref{fig:2Dstructure}). In the intermediate model, 25\% of the mass within 125~AU (disk outer radius) has $T < 20$~K. In the cold model this is 70\%, roughly consistent with the result from \citet{Kama2016} that in TW Hya $\sim$60\% of the dust mass has temperatures below 20~K. The cold model thus resembles a Class II disk. Although for the intermediate and cold model the temperature is lowered to create CO freeze-out regions, the fiducial temperature is used for the excitation calculation. This way the emission from the surface layers is not reduced through an excitation effect. If these regions would in fact also be colder, the difference with the fiducial warm model would increase.

The $^{13}$CO and C$^{18}$O emission for the three models is simulated with the 3D radiative transfer code LIME \citep{Brinch2010}. Again, the velocity inside the disk ($R <$ 125~AU) is Keplerian and the envelope material has an infalling rotating velocity profile. The inclination is set to 85$^\circ$, that is, 5$^\circ$ offset from edge-on with the disk and envelope west of the protostar tilted toward the observer (see Fig.~\ref{fig:Modelcartoon}). This is the same orientation as used by \citealt{Tobin2008} and T13, but opposite to the results presented by \citealt{Oya2015}. However, none of our conclusions are sensitive to how the inclination is defined.  Synthetic images with the same beam as the observations are then generated in CASA by using a modified version of the task \texttt{symobserve} that calculates visibilities for the observed $uv$-coverage (\texttt{symobs\_custom}, Attila Juh{\'a}sz). The continuum subtracted visibilities were then imaged to the noise level of the observations (see Table~\ref{tab:Lineparameters}). In addition, emission was simulated for a five hour integration (compared to 25 minutes for the observations) at an angular resolution similar to that of the observations. 

The by eye best matching model and lowest residuals are reached for a stellar mass of 0.4~$M_{\sun}$, although the envelope emission in the low and intermediate velocity channels is slightly overproduced for $^{13}$CO. This could be because a canonical CO abundance of $10^{-4}$ is too high for the more tenuous envelope where photodissociation may play an important role in destroying CO, unlike in the denser disk midplane where UV radiation cannot penetrate (see e.g., \citealt{Visser2009b}). Another reason could be because the large-scale cloud emission is not taken into account in the model. The effect of emission being resolved out is not as large in the simulated images as it is in the observations, that is, the simulated images still show some emission in the central most channels while the observations do not. We will therefore restrict the comparison between observations and models to velocities $\Delta v > 1.1$~km~s$^{-1}$. Lower stellar masses give lower residuals in the intermediate velocity channels, but cannot reproduce the disk emission, consistent with the results from \citet{Ohashi2014}. Since our goal is not to model the CO isotopologue emission exactly, but to focus on the effect of CO freeze-out in the disk, we show the results of the 0.4~$M_{\sun}$ models. The only difference between these models and the fiducial model presented in Sect.~\ref{sec:DiskEnvelope} is thus the stellar mass (0.4 $M_{\sun}$ instead of 0.45 $M_{\sun}$). However, the overall conclusions are valid for stellar masses between 0.19 and 0.5~$M_{\sun}$.

\begin{figure}
\centering
\includegraphics[width=0.5\textwidth]{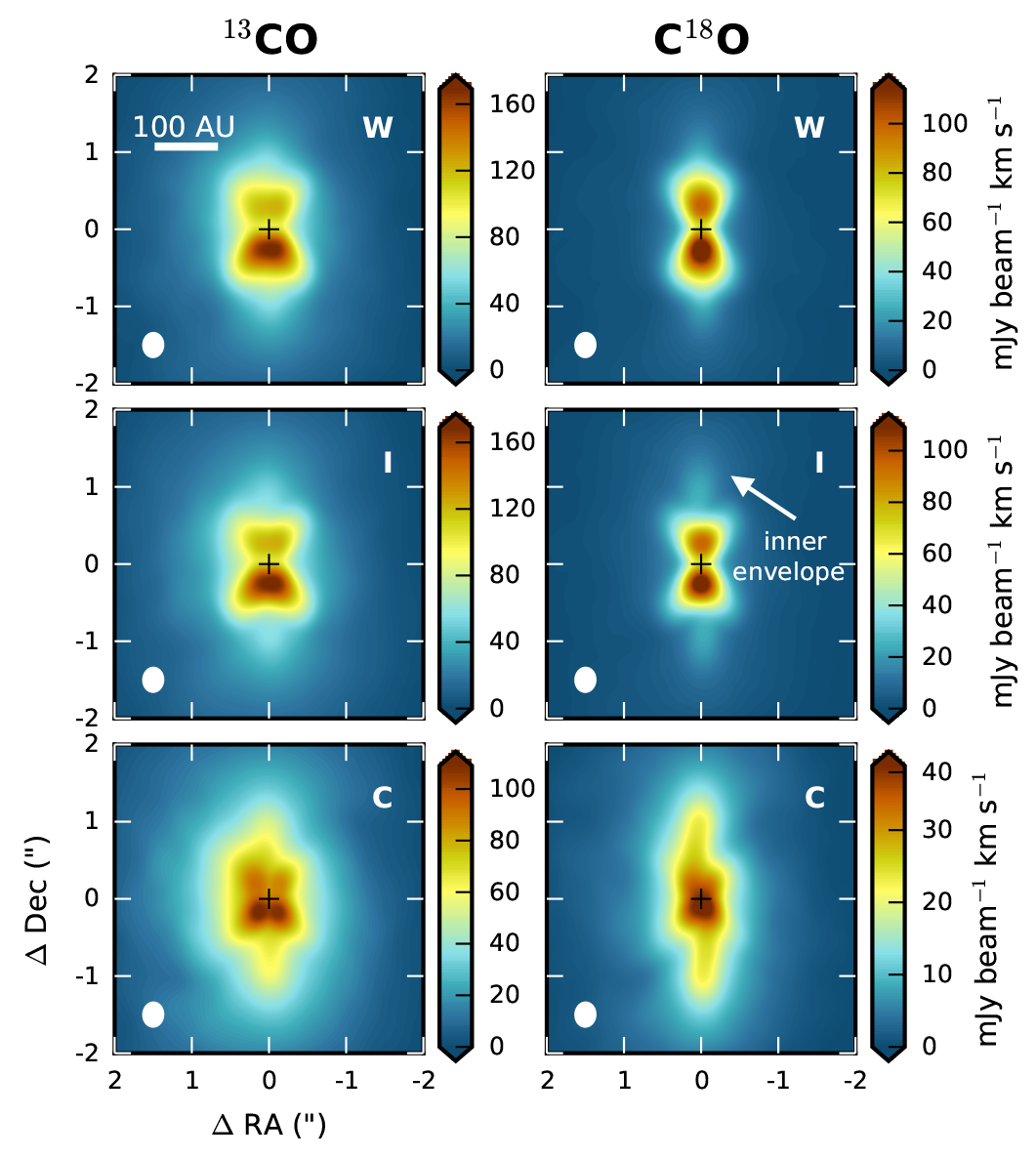}
\caption{Integrated intensity (zeroth moment) maps for $^{13}$CO (\textit{left panels}) and C$^{18}$O (\textit{right panels}) in the warm (W; \textit{top row}), intermediate (I; \textit{middle row}) and cold (C; \textit{bottom row}) model. Emission is simulated for a five hour integration using an ALMA configuration resulting in a similar beam size as the observations. The intensity (color) scale is different for each panel. The position of the continuum peak is marked by a black cross and the beam is shown in the lower left corner of the panels. } 
\label{fig:Models_M0}
\end{figure}


\begin{figure}
\centering
\includegraphics[trim={0cm 6.7cm 0cm 0.3cm},clip]{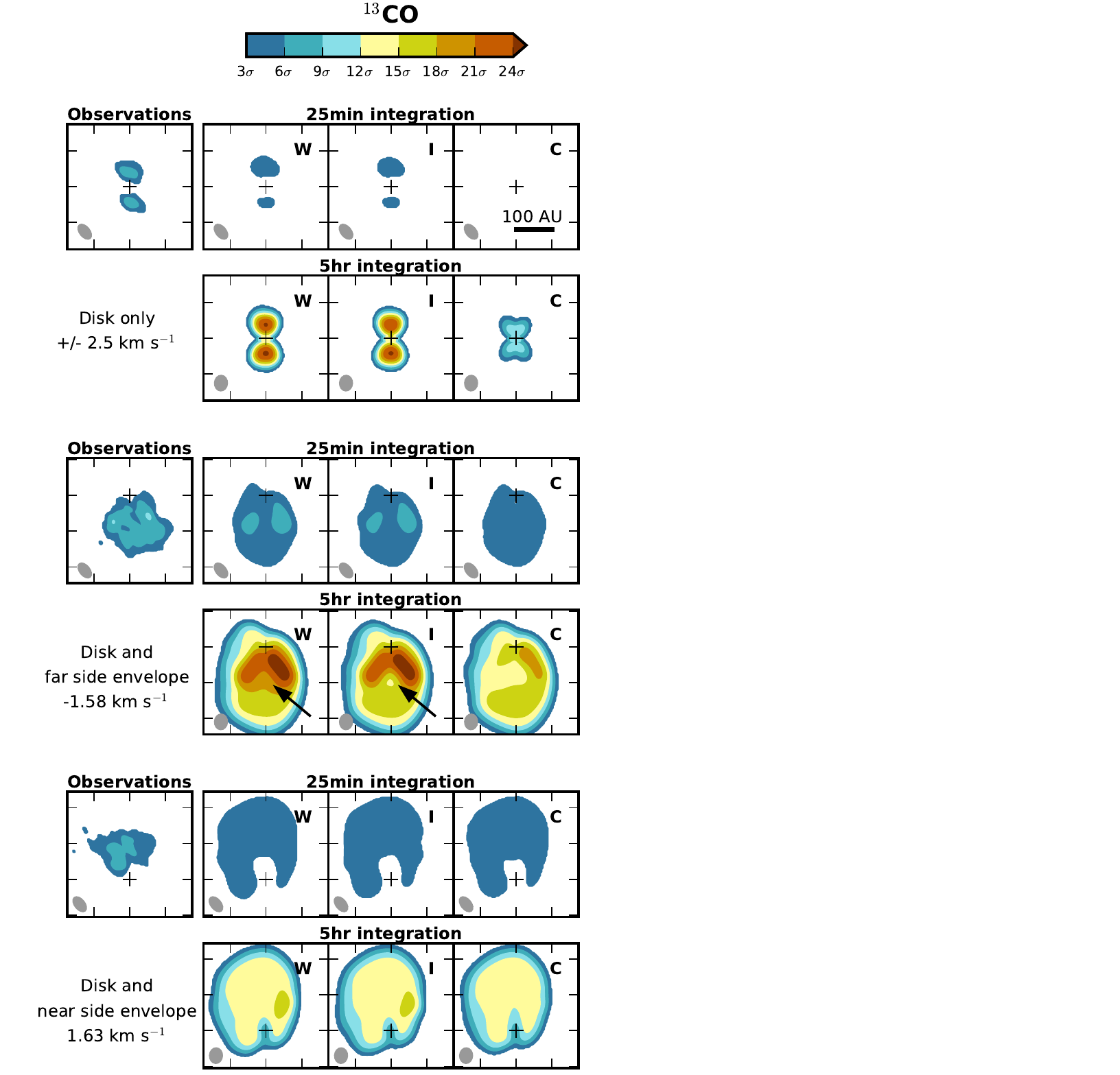}
\caption{Comparison between the $^{13}$CO observations (\textit{left column}) and the three different models (see Fig.~\ref{fig:Models}): warm (W, \textit{second column}), intermediate (I, \textit{third column}), and cold (C, \textit{fourth column}). Emission is simulated with the same visibilities as the observations (25 minutes integration) and for a longer integration time of 5 hours in a configuration producing a similar beam size. Four representative velocity channels are shown. The \textit{top panels} show emission at $\pm$2.5~km~s$^{-1}$, which is expected to originate in the disk. The \textit{middle panels} show blueshifted emission from the disk and far side of the envelope ($\Delta v = -1.58$~km~s$^{-1}$), and the \textit{bottom panels} show redshifted emission from the disk and near side of the envelope ($\Delta v = 1.63$~km~s$^{-1}$). The color scale is in steps of 3$\sigma$, which corresponds to 30.6 (8.8)~mJy~beam$^{-1}$~channel$^{-1}$ for an integration time of 25 minutes (five hours). The black arrow denotes the difference between the warm and intermediate model, which becomes clear for a five hour integration. The position of the continuum peak is marked by a cross; the vertical scale is the same for all velocities, but slightly shifted. The beam is shown in the lower left corner of each panel.}
\label{fig:Models_13CO}
\vspace{-0.2cm}
\end{figure}

\begin{figure}
\centering
\includegraphics[trim={0cm 6.5cm 0cm 0.3cm},clip]{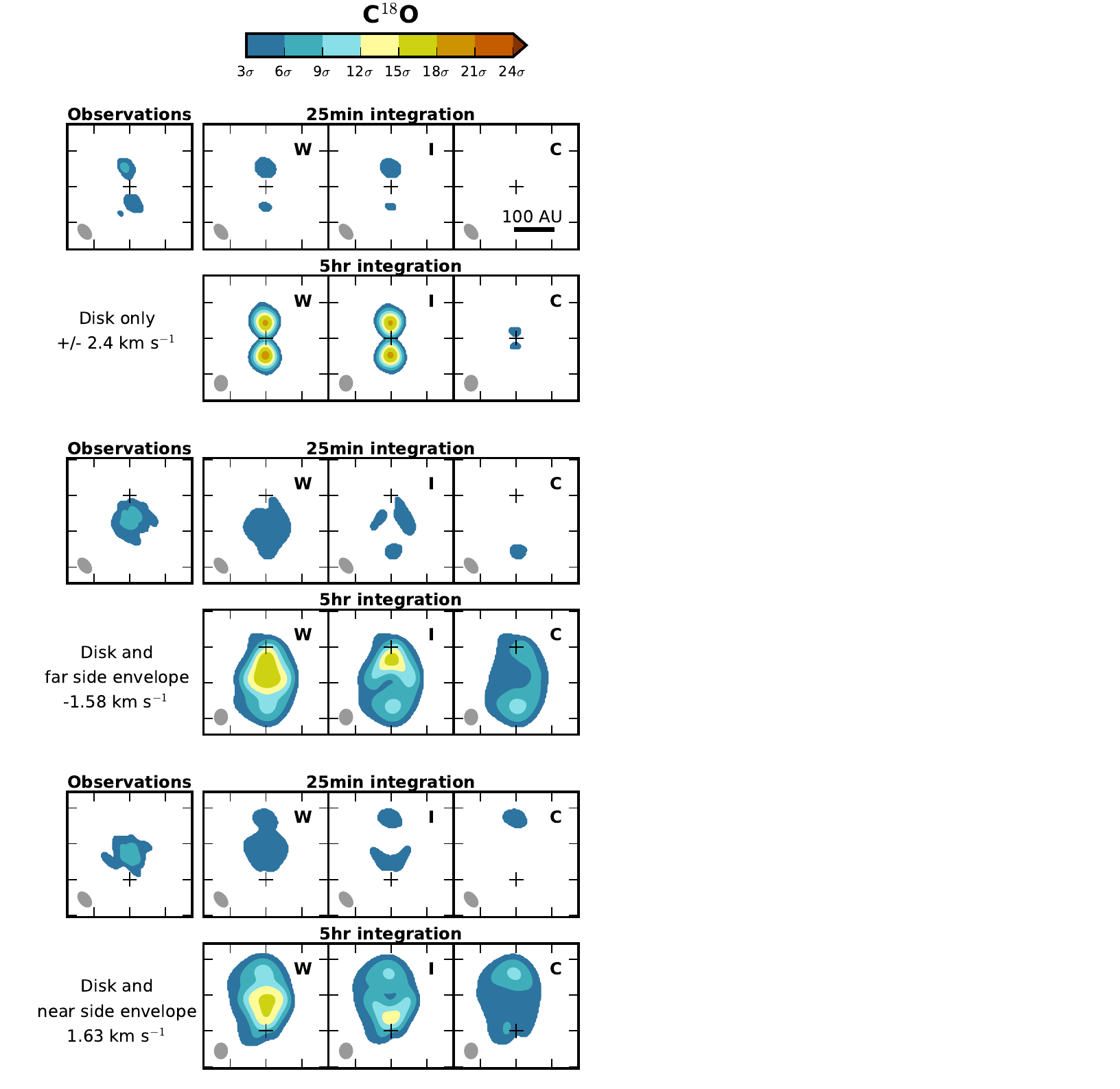}
\caption{As Fig.~\ref{fig:Models_13CO}, but for C$^{18}$O. The 3$\sigma$ level corresponds to 26.1 (7.5) ~mJy~beam$^{-1}$~channel$^{-1}$ for an integration time of 25 minutes (five hours).}
\label{fig:Models_C18O}
\end{figure}

\begin{figure}
\centering
\includegraphics[trim={0cm 11.2cm 0cm 0.6cm},clip]{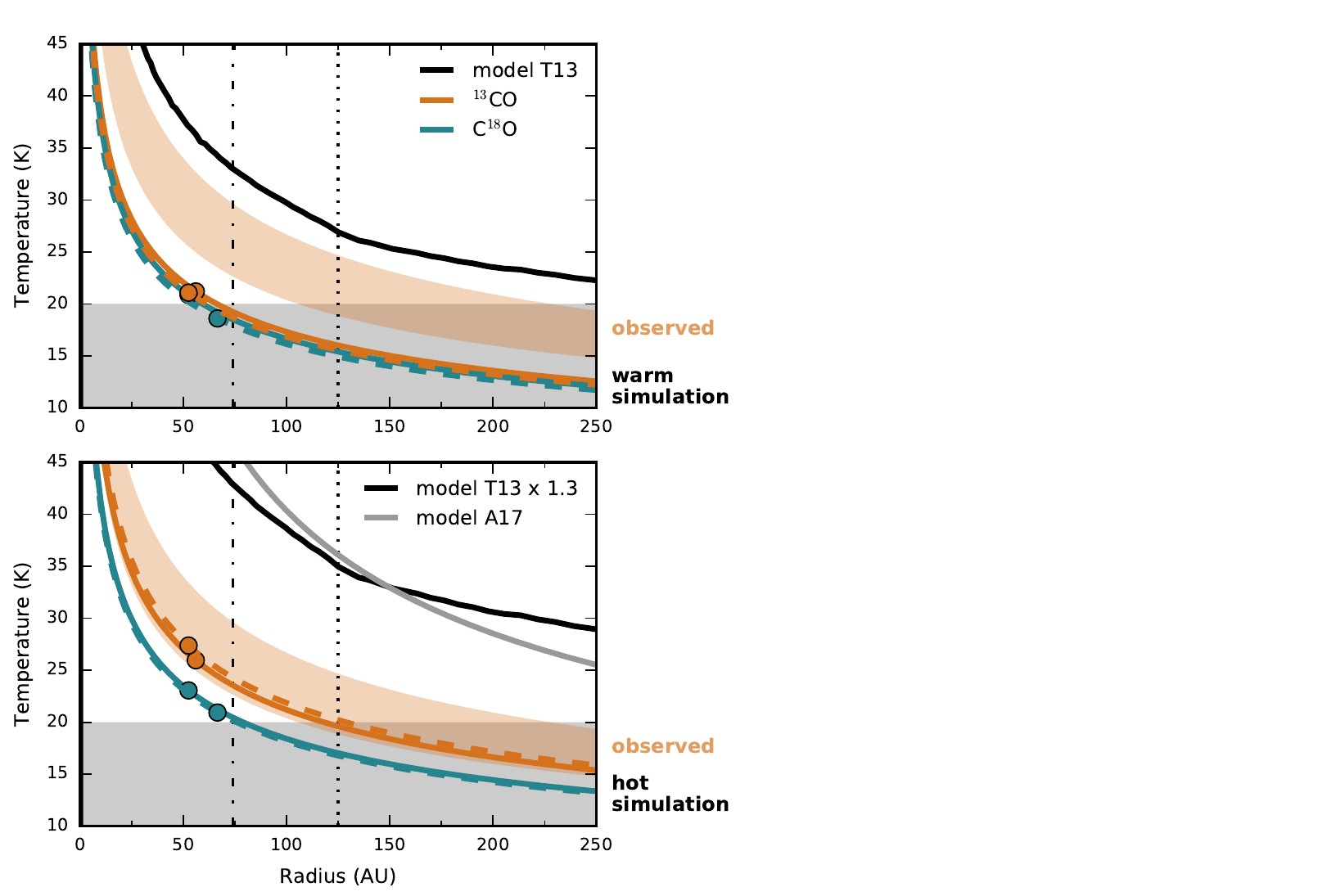}
\caption{Power law temperature profiles ($q$ = 0.35) as shown in Fig.~\ref{fig:T_profile}, but now derived from images simulated with the observed $uv$-coverage (25 minutes integration). The \textit{top panel} shows the results for the warm model, that is, the fiducial T13 temperature structure. The \textit{bottom panel} shows the results for the hot model, that is, the T13 temperature increased by 30\%. The black lines show the midplane temperature profile in the corresponding models used to generate the images. The gray line represents the midplane temperature used by A17. The shaded orange area marks the $1\sigma$ uncertainty on the temperature profile derived from the observations ($^{13}$CO, redshifted channel), as shown in Fig.~\ref{fig:T_profile}.}
\label{fig:T_profile_model}
\end{figure}


An overview of the moment zero maps for the models with five hour integrations is presented in Fig.~\ref{fig:Models_M0}. Four representative velocity channels are shown in Figs.~\ref{fig:Models_13CO} and \ref{fig:Models_C18O} for both integration times, along with the observations. The warm disk model matches the observations for both molecules quite well. The model slightly underpredicts (3--6$\sigma$ residuals) the compact emission ($<0\farcs5$), and overpredicts ($3\sigma$ residuals) the envelope emission, especially for $^{13}$CO, as discussed above. The cold model is clearly the worst match for both species, as it cannot reproduce the observed intensity. In addition, it predicts similar intensities in the disk and inner envelope (see Fig.~\ref{fig:Models_M0}), unlike what is observed (see Fig.~\ref{fig:OverviewM0}). 

With the current observations, the intermediate disk model with CO frozen out at midplane radii larger than 71~AU would be indistinguishable from the warm disk model using $^{13}$CO emission. A five hour integration would yield high enough signal to noise to distinguish both scenarios using intermediate blueshifted velocities (see Fig.~\ref{fig:Models_13CO}). However, for C$^{18}$O the warm model is a better match to the data than the intermediate model (see Fig.~\ref{fig:Models_C18O}). This difference can be explained with the C$^{18}$O isotopologue being less abundant than the $^{13}$CO isotopologue. For the blueshifted velocities (see Fig.~\ref{fig:Optdepth}, second column), the $^{13}$CO emission becomes optically thick in the inner envelope behind the disk when CO is absent in the outer disk. If the temperature in this region is similar to the temperature in the outer disk, the observed emission will hardly change. The C$^{18}$O emission on the other is expected to remain optically thin out to larger radii, probing different physical conditions deeper in the envelope. At redshifted velocities (see Fig.~\ref{fig:Optdepth}, fourth column), the $^{13}$CO emission becomes already optically thick in the inner envelope, blocking the outer disk from our view. Whether CO is present in the outer disk or not thus has no affect on the observations. The C$^{18}$O emission does not become optically thick in the envelope, so changes in the disk are visible. The effect of CO being frozen out in the outer disk is therefore likely to be stronger for the C$^{18}$O than the $^{13}$CO emission, and the less abundant C$^{18}$O seems to be a more sensitive diagnostic of the distribution of gas-phase CO in the disk. Removing CO at disk radii $>$ 71~AU does indeed significantly affect the C$^{18}$O emission, providing a worse match to the observations than the warm disk model. This is amplified for the cold disk model with CO frozen out at radii $> 23$~AU, and for five hour integrations. The intermediate and warm model can be distinguished using $^{13}$CO if the reduced temperatures, used to set the CO freeze-out region in the intermediate and cold models, are also used for the excitation calculation. In that case, the blueshifted channels (Fig.~\ref{fig:Models_13CO}, middle panels) for the intermediate model are similar to those of the cold model instead of the warm model. A disk with temperatures reduced by 40\% (w.r.t the T13 model) throughout is thus unlikely. In summary, these modeling results thus suggest that gaseous CO is likely present in the entire disk. 

The accuracy of the midplane temperature derived from the $^{13}$CO and C$^{18}$O observations can be assessed by computing a power law temperature profile from the synthetic images for the warm disk model (25 minutes integration) in the same way as done for the observations. The resulting profiles (Fig.~\ref{fig:T_profile_model}, top panel) are much colder than the input midplane temperature of the models, suggesting that the observationally derived temperature is systematically lower than the actual temperature. Moreover, these profiles are also colder than those derived from the observations. This means that the disk midplane may be warmer than the temperature profile derived by T13 from the continuum emission, which has a temperature of $\sim$26~K at the disk outer radius, because the underlying temperature profile will be substantially higher than the observationally derived one. This may be because of colder gas along the line of sight. Another effect that plays a role is beam dilution, because concatenating the high resolution data shown here with the data taken in a more compact figuration results in $\sim$10~K lower brightness temperatures ($\sim$0.1$\arcsec$ larger beam). Higher resolution observations may thus provide more accurate temperatures. 

To examine what temperature would be required to reproduce the observed brightness temperature, models were run with the T13 disk temperature increased by a certain percentage. Increasing the temperature by 30\% (hot model) results in a midplane temperature profile similar to that adopted by A17 (see Fig.~\ref{fig:T_profile_model}, bottom panel). The power law profile derived for this model from $^{13}$CO emission is within the 1$\sigma$ uncertainty for the observations, although the C$^{18}$O profiles are a few Kelvin colder (Fig.~\ref{fig:T_profile_model}). These results suggest that the L1527 disk could be warmer than 35~K.

\subsection{CO abundance}

The underabundance of gaseous CO in Class II disks, with respect to the ISM abundance of $10^{-4}$ (w.r.t. H), has been attributed to photodissociation in the upper layers of the disk and freeze-out near the outer disk midplane \citep{Dutrey1997,vanZadelhoff2001}. However, recent detailed studies of in particular TW Hya suggest that on top of this, CO is also depleted in the warm molecular layer and inside its snowline \citep{Favre2013,Nomura2016,Schwarz2016}. Moreover, a general carbon depletion was inferred from [C I] lines \citep{Kama2016}. Low CO abundances are also observed for protostellar envelopes \citep{Anderl2016}, and results from large protoplanetary disk surveys with ALMA show that this may be a common feature for Class II disks. For many disks in several star forming regions, CO (isotopologue) emission, and the subsequently derived gas masses and gas-to-dust ratios, are up to two or three orders of magnitude lower than expected \citep{Ansdell2016,Ansdell2017,Barenfeld2016,Long2017}. The main question now is whether these low CO-derived gas masses are the result of CO depletion or a sign of rapid gas loss \citep[e.g.,][]{Miotello2017}. Determining CO abundances during earlier stages of disk evolution may help solve this puzzle. 

\begin{figure*}
\centering
\includegraphics[width=\textwidth,trim={0.1cm 0.1cm 0cm 0.2cm},clip]{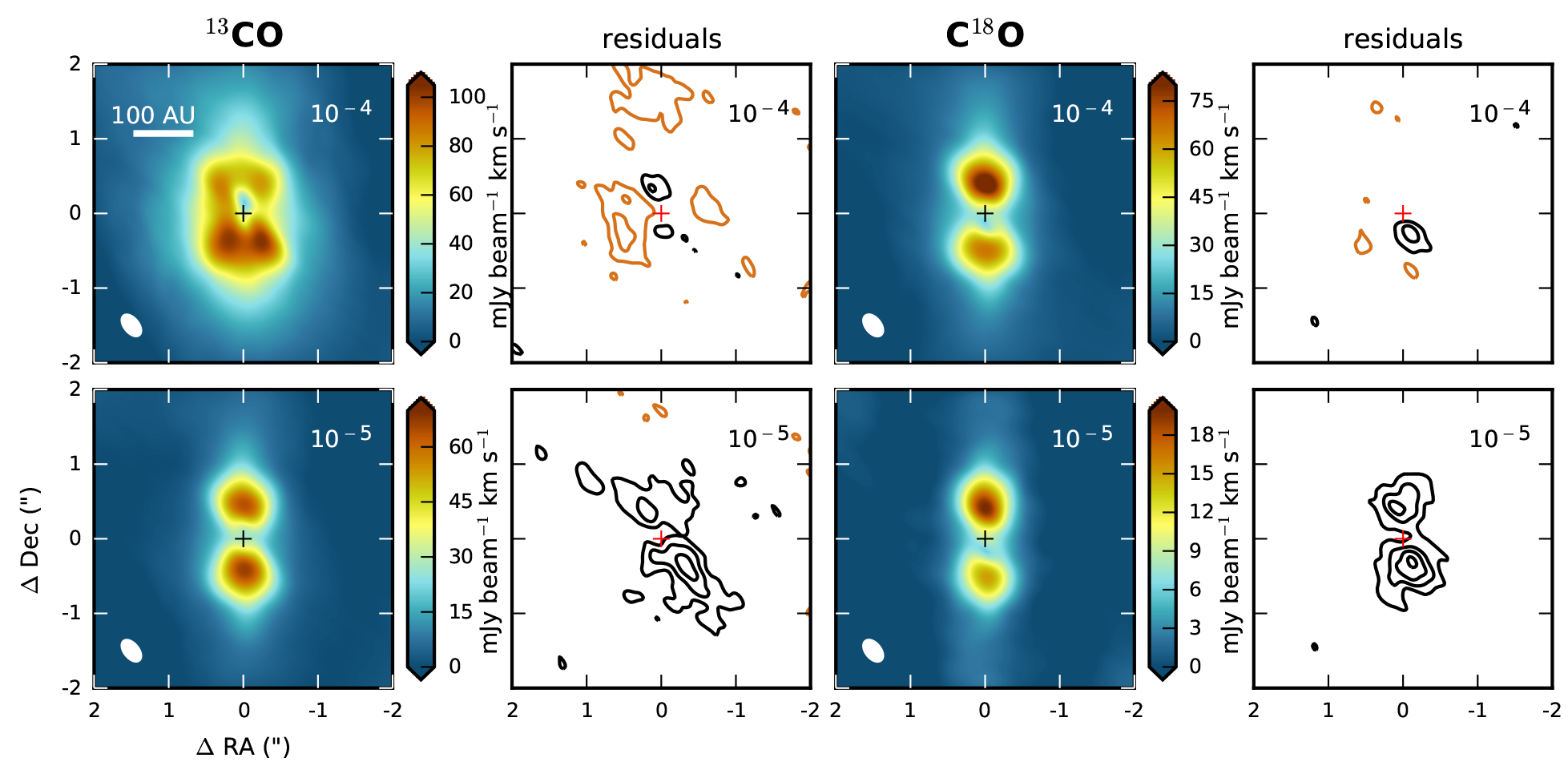}
\caption{Integrated intensity (zeroth moment) maps for $^{13}$CO (\textit{first column}) and C$^{18}$O (\textit{third column}) in the warm model with a constant CO abundance of $10^{-4}$ (\textit{top row}) and $10^{-5}$ (\textit{bottom row}). The emission is simulated with the observed $uv$-coverage (25 minutes integration). The intensity (color) scale is different for each panel. The \textit{second and fourth} columns show the residuals obtained by subtracting the models from the observations. Black contours are in steps of 3$\sigma$ starting at 3$\sigma$, while orange contours are in steps of $-3\sigma$ and start at $-3\sigma$. For $^{13}$CO (C$^{18}$O), 3$\sigma$ corresponds to 25.9 (21.8) mJy~beam$^{-1}$~km~s$^{-1}$. The position of the continuum peak is marked by a cross and the beam is shown in the lower left corner of the moment zero maps.} 
\label{fig:Models_M0_1e-5}
\end{figure*}

Given that the $^{13}$CO and C$^{18}$O emission toward L1527 is optically thick, deriving column densities and abundances is difficult. But we can invert this argument and use radiative transfer modeling to assess the CO abundance required for both molecules to be optically thick. Therefore, we made synthetic $^{13}$CO and C$^{18}$O ALMA images for the warm disk model with a constant CO abundance of $10^{-5}$, instead of $10^{-4}$ as described in Sect.~\ref{sec:Models_CO}. The resulting moment zero maps (constructed using velocities $\Delta v > 1.1$~km~s$^{-1}$) presented in Fig.~\ref{fig:Models_M0_1e-5} suggest that reducing the CO abundance by an order of magnitude would give only a $6\sigma$ detection of $^{13}$CO (peak flux of 67 mJy~beam$^{-1}$~km~s$^{-1}$), while the observed signal-to-noise ratio is $15\sigma$ (peak flux of 125 mJy~beam$^{-1}$~km~s$^{-1}$ when using the same velocity range as for the simulated images). Moreover, C$^{18}$O would only be detected at the $2\sigma$ level (peak flux of 20 mJy~beam$^{-1}$~km~s$^{-1}$), while it is observed at $14\sigma$ (peak flux of 101 mJy~beam$^{-1}$~km~s$^{-1}$). The $^{13}$CO/C$^{18}$O line ratios (not shown) are also higher than observed; generally larger than $\sim$3.5, indicating that the C$^{18}$O emission is not optically thick and even the $^{13}$CO emission becomes optically thin in some regions. A canonical CO abundance thus produces a better match to the observations than an order of magnitude reduction. 

Based on modeling of C$^{18}$O ALMA observations, A17 suggested a local enhancement of C$^{18}$O at the disk-envelope interface (that is, an ISM abundance between 80 and 88 AU), and a factor $\sim$20 lower abundance in the disk. However, none of the results presented here show an immediate need for a reduced CO abundance in the disk compared to the disk-envelope interface. On the contrary, a constant ISM abundance slightly underpredicts the emission within $\sim0\farcs5$ instead of overpredicting it. 

\subsection{N$_2$D$^+$}

N$_2$H$^+$ has been shown to trace cold ($T \lesssim$ 20~K) and dense \mbox{($n > 10^{5}$~cm$^{-3}$)} prestellar and protostellar environments where CO is frozen out \citep[e.g.,][]{Bergin2001,Caselli2002a,Jorgensen2004}. Another chemical effect of low temperatures and CO freeze-out is the enhancement of deuterated molecules, such as N$_2$D$^+$ \citep[e.g.,][]{Caselli2002b,Crapsi2005,Roberts2007}. With the resolution provided by ALMA, N$_2$H$^+$ emission has been spatially resolved in a few protoplanetary disks, where it was shown to originate outside the CO snowline \citep{Qi2013,Qi2015}. Recently, N$_2$D$^+$ has also been observed toward one T Tauri and one Herbig disk \citep[][respectively]{Huang2015,Salinas2017}. In both cases the measured N$_2$D$^+$/N$_2$H$^+$ ratio of 0.3--0.5 matches  the highest values found for protostellar envelopes \citep{Emprechtinger2009,Tobin2013b}. The latter single dish studies find one of the lowest N$_2$D$^+$/N$_2$H$^+$ ratios for L1527 ($<$ 0.001) in samples of $\sim$20 protostars. 

So far, no observations of N$_2$H$^+$ have been carried out for the L1527 disk. Based on a small chemical network for N$_2$H$^+$ \citep{vantHoff2017}, abundances of $\sim$10$^{-10}$ are expected for the densities in the T13 model if CO freeze-out would occur. This chemical model incorporates the main formation and destruction processes for N$_2$H$^+$, as well as freeze-out, thermal desorption and photodissociation of CO and N$_2$. An N$_2$H$^+$ abundance of $\sim$10$^{-10}$ is consistent with predictions from full chemical models \citep[e.g.,][]{Walsh2012,Aikawa2015}. If the N$_2$D$^+$/N$_2$H$^+$ ratio in the disk were similar to the value in the envelope (i.e., $<$ 0.001), an N$_2$D$^+$ abundance $\lesssim 10^{-13}$ (w.r.t. H) would be expected in the disk. To assess whether the observations are sensitive enough to have detected N$_2$D$^+$ abundances $\lesssim 10^{-13}$, synthetic images were created for the intermediate and cold disk model. In these models, N$_2$D$^+$ is present at constant abundance throughout the CO freeze-out region. Based on these simulations, only abundances $\gtrsim10^{-11}$ could have been detected with the observations presented here. These findings can also be phrased in terms of column density. For the intermediate model, that is, N$_2$D$^+$ present outside 71~AU, the N$_2$D$^+$ column density along the midplane for an abundance of $10^{-13}$ is $3\times10^{10}$~cm$^{-2}$. This is two orders of magnitude lower than the upper limit derived from the observations ($\sim 2-4\times10^{12}$~cm$^{-2}$). Even for the cold model with N$_2$D$^+$ at radii $> 23$~AU, which cannot reproduce the CO observations, the N$_2$D$^+$ column density would still be an order of magnitude below the observed upper limit. The observations thus seem not sensitive enough to completely rule out the presence of N$_2$D$^+$ in the outer disk.

However, \citet{Emprechtinger2009} found that the lowest N$_2$D$^+$/N$_2$H$^+$ ratios were seen at the highest dust temperatures. In particular, the low ratio for L1527 suggests dust temperatures above $\sim$25~K on envelope scale. If the deuterium fractionation in the disk is similar to that in the envelope, this would then be in agreement with a warm disk.   

\section{Discussion} \label{sec:Discussion}

Based on analysis and modeling of the optically thick $^{13}$CO and C$^{18}$O emission observed with ALMA, the young disk around L1527 is likely to be warm enough to prevent CO from freezing out ($T \gtrsim$ 20--25~K). This is in contrast with observations of more evolved Class~II disks which are shown to have relatively large cold outer regions where CO is frozen onto dust grains. 

The main difficulty in characterizing the physical properties of young disks is disentangling disk emission from emission originating in the envelope. As shown in Sect.~\ref{sec:DiskEnvelope}, a rough division can be made kinematically, but it becomes more complicated if the emission is optically thick. In the latter case, the angular offset at which emission is observed no longer correlates to the physical radius at which the radiation is emitted. Therefore, careful analysis is required when characterizing disks that are still embedded in their envelope.

Although an accurate determination of the origin of the optically thick $^{13}$CO and C$^{18}$O emission requires detailed knowledge of the systems physical structure, our inferred power law temperature profile does not depend on these parameters. In the analysis, the midplane velocity component along the line of sight is used to predict the velocities that contain only disk emission. This velocity component depends on the stellar mass, disk radius and envelope velocity profile. For both the T13 model ($M_{\star} = 0.19 M_{\sun}$, $R_{\rm{disk}} = 125$ AU, and infalling rotating envelope) and the A17 model ($M_{\star} = 0.45 M_{\sun}$, $R_{\rm{disk}} = 74$, and a reduced free fall velocity in the envelope), emission at $\vert\Delta v \vert \geq$ 2.6~km~s$^{-1}$ is expected to originate only in the disk. The infalling rotating velocity profile predicts higher envelope velocities than the reduced free fall profile, but even for a central mass of 0.45 $M_{\sun}$ no envelope contamination is expected at $\vert\Delta v \vert \geq$ 2.6~km~s$^{-1}$. So independent of the assumed model, the brightness temperature at $\vert\Delta v \vert \geq$ 2.6~km~s$^{-1}$ is likely to probe the temperature in the disk. 

\subsection{Dust opacity} \label{sec:Disc_model}

\begin{figure}
\centering
\includegraphics[width=0.48\textwidth,trim={0cm 0.5cm 0cm 0cm},clip]{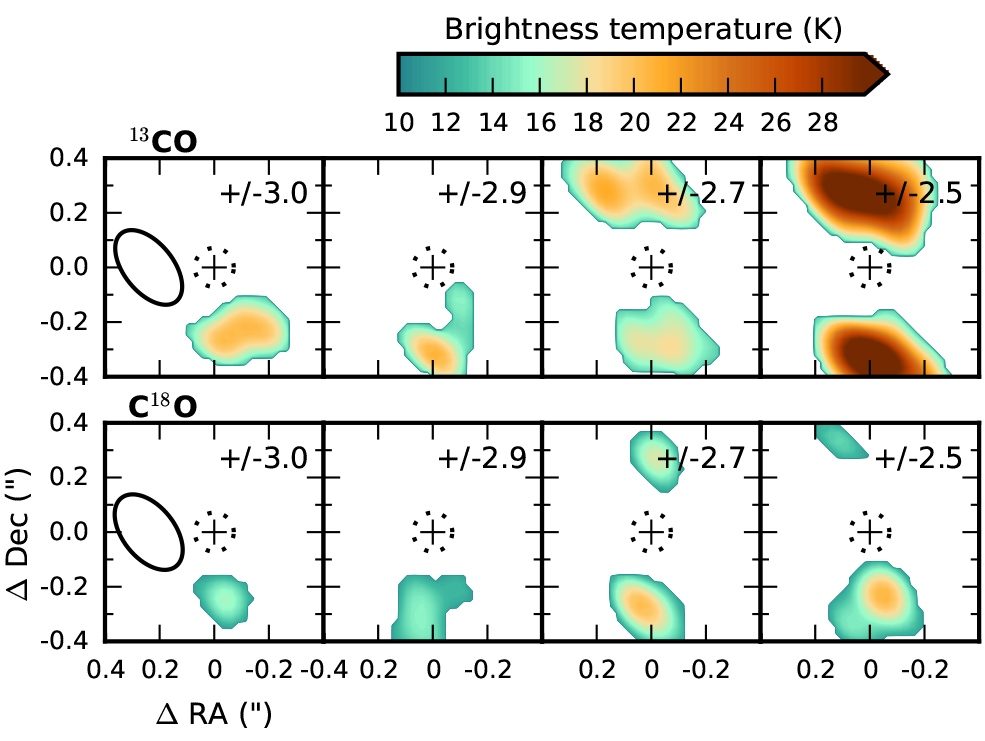}
\caption{Zoom in on the highest velocity channels for $^{13}$CO (\textit{top panels}) and C$^{18}$O (\textit{bottom panels}), highlighting that the emission centroid does not move closer to the continuum peak position (marked with a cross) with increasing velocity offset. Only emission above the 3$\sigma$ level is shown. The velocity relative to the systemic velocity of $v_{\rm{lsr}} = 5.9$~km~s$^{-1}$ is listed in the top right corner of each panel. For each velocity offset, both blue- (south of source position) and redshifted emission (north of source position) are shown in the same panel. The dotted contour marks the inner 10~AU where the T13 model predicts the continuum to be optically thick. The black contour in the left panels shows the synthesized beam.}
\label{fig:ChannelsZoomin}
\end{figure}

To study whether the observations can distinguish between a warm and cold disk, we use the best-fit model from T13, which was obtained by fitting 3D radiative transfer models to sub/millimeter dust emission and infrared scattered light. As pointed out by the authors, the most robust parameters are the disk radius and vertical height. Meanwhile, the disk mass and radial density profile are the most degenerate parameters. Both depend on assumptions made for the dust opacity. Interesting in this respect is that the observations of the CO isotopologues presented here hint at the 1.3 mm continuum being optically thick close to the protostar. As can be seen in Fig.~\ref{fig:ChannelsZoomin}, the centroids of both the $^{13}$CO and C$^{18}$O emission remain offset from the stellar position even at the highest velocities. This can be caused by the dust being optically thick in this inner region, obscuring the CO emission. For the T13 density structure and dust opacity ($\kappa =$3.1~cm$^2$~g$^{-1}$ at 200 GHz), the continuum becomes optically thick in the inner $\sim$10~AU. If the dust is indeed optically thick in the inner disk, the dust mass derived by T13 may be underestimated by a factor of a few. This could lead to our modeled CO isotopologue emission being more optically thin than in the observations, and therefore modeled brightness temperature to be slightly lower than observed. Modeling of higher resolution multi-wavelength observations is needed to better constrain the dust opacity.

\subsection{Comparison with other observations}

\subsubsection{L1527}

A good consistency check is to determine whether a warm disk is consistent with the observed luminosity and mass accretion rate. For L1527, mass accretion rates between $\sim 6.6\times10^{-7} M_{\sun}$~yr$^{-1}$ \citep[][for a 0.19 $M_{\sun}$ star assuming $L_{acc} = GM\dot M/R_{\star}$]{Tobin2012} and $\sim 1.8\times10^{-6} M_{\sun}$~yr$^{-1}$ \citep[][for a 0.1 $M_{\sun}$ star]{Ohashi1997} have been estimated. Scaling the value derived by \citet{Tobin2012} for a 0.45 $M_{\sun}$ star yields a rate of $\sim2.9\times10^{-7} M_{\sun}$~yr$^{-1}$. According to physical models by \citet{Harsono2015}, no CO freeze-out is expected in an embedded disk for accretion rates $> 10^{-7}M_{\sun}$~yr$^{-1}$. In addition, \citet{Frimann2017} present an expression to predict the size of the C$^{18}$O-emitting region based on a set of axi-symmetric radiative transfer models for protostellar envelopes: 
\begin{equation}
\mathrm{FWHM} = a \left( \frac{d}{235\mathrm{pc}} \right)^{-1} \left( \frac{L_{\mathrm{bol}}}{1 L_{\sun}} \right)^{0.41},
\end{equation} where $d$ is the distance to the source, and the angular coefficient $a$ equals diameters of 0.89$\arcsec$ or 1.64$\arcsec$ for a CO freeze-out temperature of 28 or 21~K, respectively. For a bolometric luminosity of 1.9~$L_{\sun}$ \citep{Kristensen2012}, the FWHM of the C$^{18}$O-emitting region is then expected to be 1.94$\arcsec$ or 3.58$\arcsec$ for the respective freeze-out temperatures. This is larger than the derived disk diameters of 1.06$\arcsec$ (A17) or 1.79$\arcsec$ (T13). These models thus suggest that a disk warm enough to retain CO in the gas-phase is consistent with the measured accretion rates and luminosity. 

\citet{Sakai2014a} assess the physical conditions in L1527 by using a non-LTE large-velocity-gradient (LVG) code. For $1\arcsec \times 1\arcsec$ regions the kinetic temperature of the  cyclic-C$_3$H$_2$ emitting region is found to be between 23 and 33~K. This is consistent with the temperature profile from T13. In the SO emitting region around the postulated centrifugal barrier ($\sim$100~AU), the gas temperature is suggested to be higher than 60~K. Recent analysis of higher angular resolution SO observations suggest even higher temperatures ($\sim$200~K) around the centrifugal barrier \citep{Sakai2017}. Due to a low signal-to-noise ratio within 100~AU, the authors cannot derive a temperature in the disk. The highest velocity offset where SO is present is $|\Delta v| \sim$ 1.8~km~s$^{-1}$. At these blueshifted velocities the outer edge of the disk and inner region of the envelope are expected to be seen. Although the temperature in the outer disk cannot be determined without knowledge of the disk outer radius, a local temperature enhancement at the disk-envelope interface is expected to be visible. However, the $^{13}$CO and C$^{18}$O observations presented here show no evidence of such high temperatures, despite being present in the gas phase around the centrifugal barrier. In addition, the beam size for the CO isotopologue observations is comparable to or much smaller than the beams for SO \citep[][resp.]{Sakai2017,Sakai2014a}, so the lower temperatures based on CO are not likely due to beam dilution.

\subsubsection{Other Class 0 and Class I disks}

Detailed studies of the gas temperature have not been performed for other young disks. \citet{Murillo2015} show that DCO$^+$ emission toward VLA1623 surrounds the disk traced in C$^{18}$O emission. Since formation of DCO$^+$ is enhanced at temperatures $\lesssim$30~K, this suggests that the temperature in the disk is too high for CO freeze-out. Attempts to assess whether young disks generally lack CO ice may be based on luminosity or the physical structure derived for the envelope. However, both approaches would yield large uncertainties. In the former case because the temperature structure does not only depend on luminosity, but also on the disk mass. In addition, whether the CO freeze-out temperature will be reached in the disk will depend on disk radius. The latter method is not likely to give reliable disk structures as the temperature and density in the inner envelope not necessarily continue smoothly into the disk \citep{Persson2016}. For example, disk shadowing may lower the temperature in the envelope just outside the disk \citep{Murillo2015}. High resolution (line) observations are thus required to derive the temperature in young disks. 

The edge-on configuration of L1527 allows for the midplane temperature to be derived from $^{13}$CO and C$^{18}$O emission. For less inclined disks, this will be much harder, if not impossible, since the emission is likely to become optically thick in the inner envelope or disk surface layers. Even less abundant isotopologues as C$^{17}$O or $^{13}$C$^{18}$O may be used \citep{Zhang2017}, but likely require long integration times. The easiest way to determine the presence of absence of cold CO-depleted gas may therefore be through observations of cold gas tracers such as N$_2$H$^+$,  N$_2$D$^+$ or DCO$^+$.

\subsubsection{Class II disks}

Until recently, thermal structures for Class II disks were derived from radiative transfer modeling of continuum emission, analysis of multiple CO lines \citep[e.g.,][]{Dartois2003,Schwarz2016}, or observations of chemical tracers of CO freeze-out \citep{Qi2013,Qi2015,Mathews2013}. These results are generally model-dependent, or limited to the vertical region where the emission originates. With the high spatial and spectral resolution and sensitivity provided by ALMA it is now possible to directly map the 2D temperature structure. \citet{Dutrey2017} present a method to derive the temperature structure (both radially and vertically) of near edge-on disks using position-velocity diagrams. They apply this technique to CO observations of the Flying Saucer disk and find that the overall thermal structure is consistent with a cold ($<15-12$~K) midplane where CO is frozen out and warmer upper layers. For intermediate inclined disks, emission from different regions can be separated due to the Keplerian rotation, given that the spectral resolution is high enough. \citet{Pinte2018} use this strategy to measure the altitude, velocity and temperature of the CO emitting layers in the disk around IM Lupi. Their results are again consistent with a CO-depleted midplane. 

Recent surveys of Class II disks have revealed gas-to-dust ratios upto two or three orders of magnitude below the ISM value of 100 \citep{Ansdell2016,Ansdell2017,Barenfeld2016,Long2017}. Whether this is the result of general gas loss or a depletion of CO on top of freeze-out and photodissociation remains an open question. Detailed studies of TW Hya, one of the few disks for which the gas mass is derived from HD emission, point to an additional depletion of CO \citep{Favre2013,Schwarz2016}, and carbon in general \citep{Kama2016}. Possible explanations for these low abundances are chemical evolution into more complex species, either in the gas or in the ice, or the lock up of carbon in large icy bodies. Determination of the CO abundance in younger disks may help shed some light on this problem. For example, low CO abundances in embedded disks would make general gas loss a less likely explanation. However, at least for L1527, we find no evidence for an order of magnitude reduction in CO abundance. This suggests that L1527 has not gone through a long dense pre-stellar core phase in which CO has been turned into other species, and hints that the physical or chemical processes that are responsible for observed low CO abundances occur on timescales longer than the Class 0 lifetime, or start at later stages. 

\subsection{Implications of a warm disk in the embedded phase}

An important question in planet formation theory is whether the composition of the planet forming material is inherited from the parental cloud, processed on the way to the disk, or completely reset to atoms, as this will influence planet composition \citep[e.g.,][]{Aikawa1999,Hincelin2013,Pontoppidan2014,Eistrup2016}. Physicochemical models by \citet{Visser2009a} have shown that not all ices are directly incorporated into the disk, but can instead have desorbed and even recondensed. In addition, \citet{Drozdovskaya2016} show that processing of ices while the material falls onto the disk can lead to enhancement of CO$_2$ ice and organics, at the expense of CO and methanol ice, respectively. The observed absence of a CO freeze-out region in the L1527 disk thus confirms that the chemical composition of the disk may not be entirely inherited from the dense cloud core, where most CO is frozen out.

In particular, the C/O ratio in the ice will thus not be inherited. The bulk of oxygen is expected to be in H$_2$O and CO$_2$, while CO is a dominant carbon carrier. Therefore, desorption of CO will relatively deplete the ice of carbon, lowering the C/O ratio \citep{Oberg2011}. The C/O ratio of icy grains in a warm disk will thus be lower than in the parental cloud. Furthermore, if planetesimals are formed during the embedded stage, these planetesimals will have lower bulk C/O ratios than their counterparts formed in the cold outer regions of Class II disks. If young disks in general are found to be warm, an interesting question would be how long this warm phase lasts. That is, do disks remain warm throughout the Class I phase, or have they cooled sufficiently before the envelope is fully dissipated? Answering this question requires detailed studies of young disks during both the Class 0 and Class I phase.


\section{Conclusions}  \label{sec:Conclusions}

Temperature is an important parameter for the physical and chemical evolution of disks around young stars, and therefore for the outcome of planet formation. Both observations and models point to Class II disks having a large cold gas reservoir where CO is frozen out. In contrast, models suggest that during earlier stages the disk can be warm enough to prevent CO freeze-out. However, so far the temperature structure of embedded disks remained poorly constrained observationally. In this work we have analyzed archival ALMA data of $^{13}$CO, C$^{18}$O and N$_2$D$^+$ to elucidate the midplane temperature structure of the edge-on Class 0/I disk around L1527. Based on the following results we conclude that this young embedded disk is indeed warm enough ($\gtrsim$20--25~K) to have CO in the gas-phase throughout the entire disk: 
\begin{itemize}
	\item $^{13}$CO and C$^{18}$O emission is detected throughout the disk and inner envelope, while N$_2$D$^+$, which can only be abundant when CO is frozen out, is not detected. 
	\item Deriving the exact radius at which the $^{13}$CO and C$^{18}$O emission originates is non-trivial due to CO being present in both the disk and inner envelope, and the emission being optically thick. However, the channel maps provide a global view of the temperature in the disk and inner envelope, and show temperatures above $\sim$25~K along the midplane. 
	\item A radial power law for the midplane temperature extrapolated from the highest velocity channels suggests that the disk is warm ($>$20~K) out to at least $\sim$75~AU. Deriving such power law from synthesized images indicates that these observations underestimate the midplane temperature. More accurate temperature measurements are expected for higher resolution observations. Moreover, the temperature derived from the synthesized images is lower than that from the observations. The disk midplane may thus be warmer than the temperature profile derived by T13 from the continuum emission, which has a temperature of $\sim$26~K at the disk outer radius (125 AU). 
	\item Radiative transfer modeling for a warm (no CO freeze-out), intermediate (CO snowline at 71~AU) and cold disk (CO snowline at 23~AU) suggest that the less abundant C$^{18}$O is a more sensitive diagnostic of the CO distribution in the disk than $^{13}$CO. The warm model best reproduces the observations for both molecules, while the intermediate model is a worse match for C$^{18}$O. 
\end{itemize}
In addition, based on radiative transfer modeling using the best-fit model from T13, there is no evidence for a low gas-phase CO abundance ($\ll 10^{-4}$), as has been suggested for some Class II disks, or an increased abundance at the disk-envelope interface, as suggested by A17. 

Conclusions about whether young disks in general are warm, and if this is the case, how long this warm phase lasts, will require detailed studies of more disks in both the Class~0 and Class~I phase.


\begin{acknowledgements} 
We would like to thank the referee for constructive and insightful comments. This paper makes use of the following ALMA data: ADS/JAO.ALMA\#2013.1.01086.S. ALMA is a partnership of ESO (representing its member states), NSF (USA) and NINS (Japan), together with NRC (Canada), MOST and ASIAA (Taiwan), and KASI (Republic of Korea), in cooperation with the Republic of Chile. The Joint ALMA Observatory is operated by ESO, AUI/NRAO and NAOJ. Astrochemistry in Leiden is supported by the European Union A-ERC grant 291141 CHEMPLAN, by the Netherlands Research School for Astronomy (NOVA) and by a Royal Netherlands Academy of Arts and Sciences (KNAW) professor prize. M.L.R.H acknowledges support from a Huygens fellowship from Leiden University, and J.J.T acknowledges support from grant 639.041.439 from the Netherlands Organisation for Scientific Research (NWO). 
\end{acknowledgements}


\bibliographystyle{aa} 
\bibliography{References}


\begin{appendix}


\section{$^{13}$CO/C$^{18}$O line ratio} \label{ap:Lineratio}

\begin{figure*}
\centering
\includegraphics[width=\textwidth,trim={0cm 16cm 0cm 0.0cm},clip]{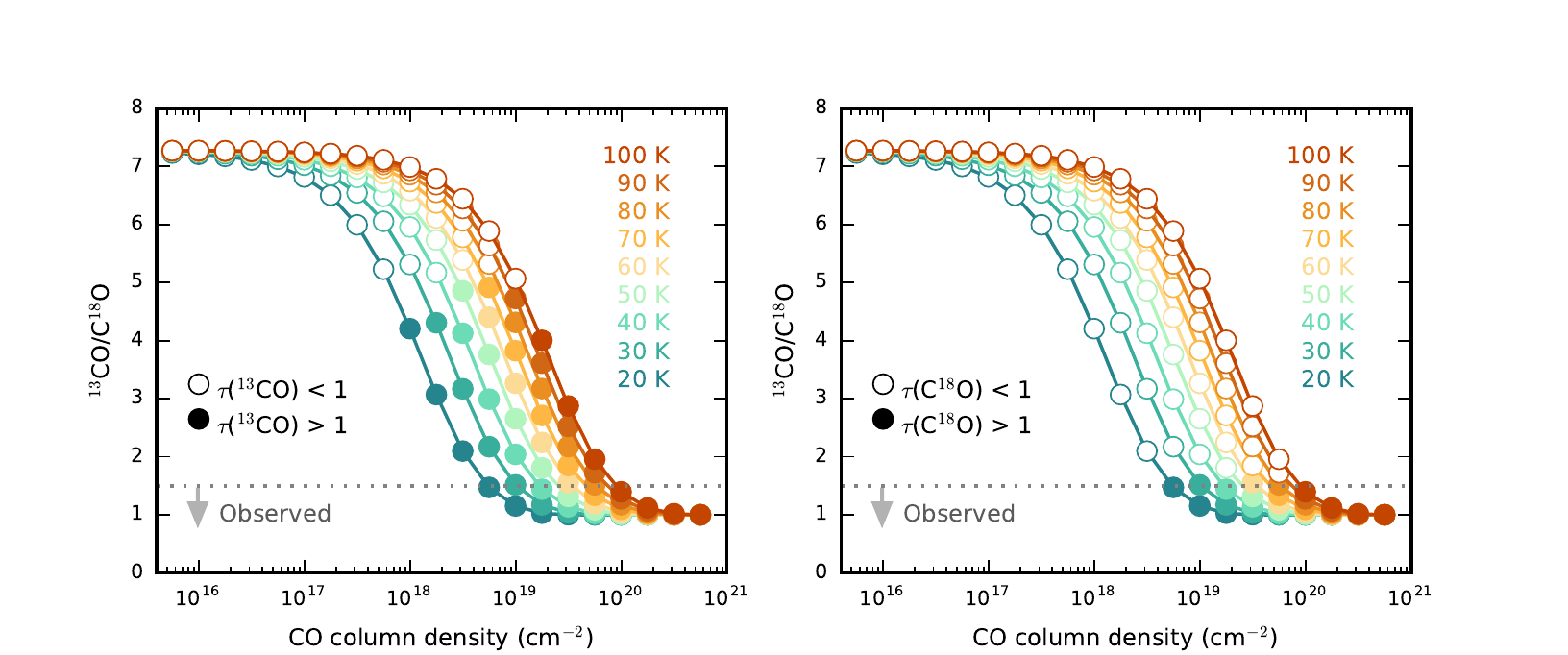}
\caption{$^{13}$CO ($J=2-1$)/C$^{18}$O ($J=2-1$) line intensity ratio calculated with RADEX for a range of temperatures (different colors) and CO column densities. Isotope ratios of [$^{12}$C]/[$^{13}$C] = 77 and [$^{16}$O]/[$^{18}$O] = 560 are adopted. In the \textit{left panel} (\textit{right panel}), open circles indicate that the $^{13}$CO (C$^{18}$O) emission is optically thin (i.e., $\tau < 1$), whereas filled circles are used when the $^{13}$CO (C$^{18}$O) emission is optically thick (i.e., $\tau > 1$). Values below the dotted line are observed toward the  L1527 midplane.}
\label{fig:RADEXLineratio}
\end{figure*}

Figure ~\ref{fig:RADEXLineratio} shows the $^{13}$CO/C$^{18}$O line ratios from RADEX calculations for different temperatures and CO column densities. 

\section{Brightness temperature} \label{ap:Tb}

\begin{figure*}
\centering
\includegraphics[width=\textwidth,trim={0cm 12.5cm 0cm 0.8cm},clip]{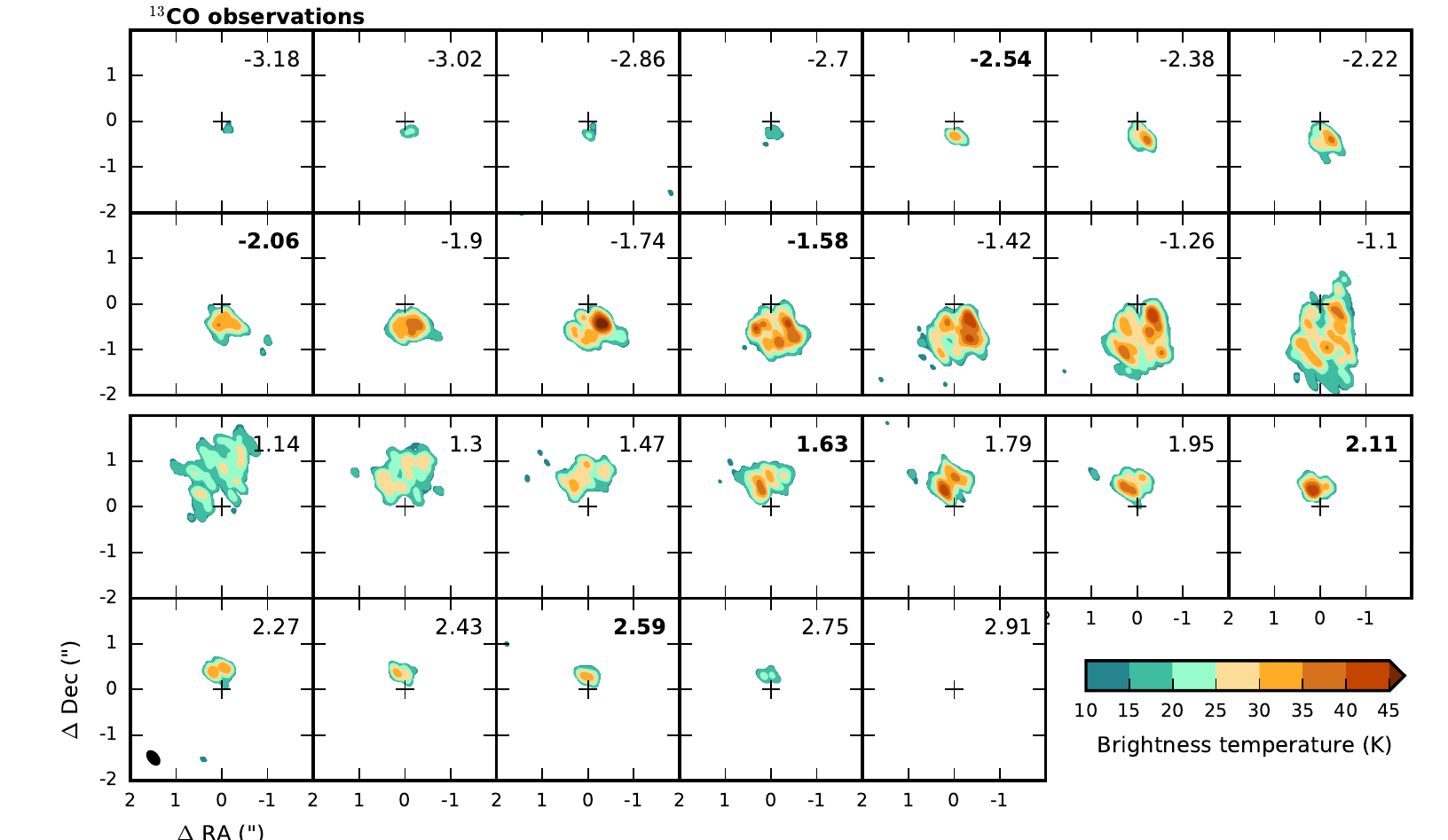}
\caption{Observed brightness temperature for the $^{13}$CO $J=2-1$ transition, as shown for six channels in Fig.~\ref{fig:TB}, but now for all channels (expect the central channels, $\pm \sim1.0$~km~s$^{-1}$, in which most or all emission is resolved out). Channel velocities with respect to the systemic velocity of $v_{\rm{lsr}} = 5.9$ km s$^{-1}$ are listed in the top right corner of each panel. Velocities presented in Fig.~\ref{fig:TB} are highlighted in boldface.}
\label{fig:13CO_TB_all}

\centering
\includegraphics[width=\textwidth,trim={0cm 12.5cm 0cm 0.5cm},clip]{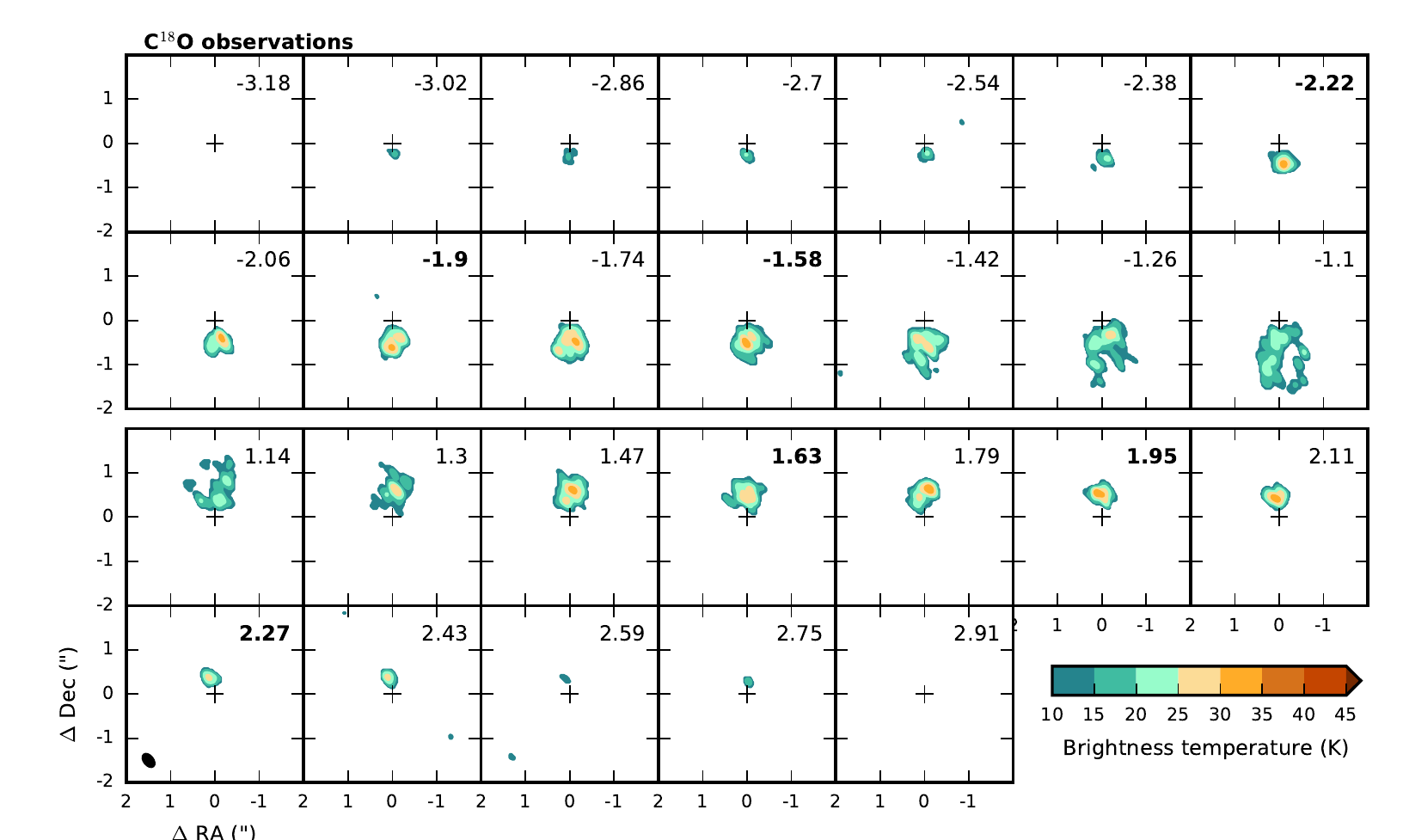}
\caption{As Fig.~\ref{fig:13CO_TB_all}, but for the C$^{18}$O $J=2-1$ transition.}
\label{fig:C18O_TB_all}
\end{figure*}

The observed brightness temperature for $^{13}$CO and C$^{18}$O (as shown in Fig.~\ref{fig:TB}) for all channels are presented in Figs.~\ref{fig:13CO_TB_all} and \ref{fig:C18O_TB_all}, respectively. The central channels in which most emission is being resolved out are omitted.

\section{Physical model for disk and envelope} \label{ap:Model}

To analyse the observed CO isotopologue emission and make synthetic images, we adopt for the physical structure the best fit model from T13. This model is the result of fitting a large grid of 3D radiative transfer models to the thermal dust emission in the sub/millimeter, the scattered light $L^{\prime}$ image, and the multi-wavelength SED. A full description of the modeling can be found in \citet{Whitney2003}, \citet{Tobin2008} and T13, while a brief summary is given below. 

The model consists of a Keplerian disk embedded within an infalling rotating envelope. For the envelope, the UCM model is adopted \citep{Ulrich1976,Cassen1981}, which describes the collapse of an isothermal, spherically symmetric, and uniformly rotating cloud. It assumes free fall with each particle conserving angular momentum. As a result, particles move along parabolic paths, with the velocity at point $(r,\theta)$ given by:
\begin{equation} \label{eq:UCM1}
v_r(r,\theta) = - \left( \frac{GM_\star}{r} \right) ^ {1/2} \left( 1 + \frac{\cos\theta}{\cos\theta_0} \right) ^ {1/2},
\end{equation}
\begin{equation}
v_{\theta}(r,\theta) = \left( \frac{GM_\star}{r} \right) ^ {1/2} (\cos\theta_0 - \cos\theta) \left( \frac{\cos\theta_0 + \cos\theta}{\cos\theta_0\sin^2\theta} \right) ^ {1/2}, 
\end{equation}
\begin{equation} \label{eq:UCM2}
v_{\phi}(r,\theta) =  \left( \frac{GM_\star}{r} \right) ^ {1/2} \frac{\sin\theta_0}{\sin\theta} \left( 1 - \frac{\cos\theta}{\cos\theta_0}  \right) ^{1/2}, 
\end{equation}
where $\theta_0$ is the inclination of the particle's orbital plane relative to the rotation axis. $\theta_0$ can be found analytically by solving the following equation:
\begin{equation}
r = R_c \frac{\cos\theta_0\sin^2\theta_0}{\cos\theta_0 - \cos\theta}, 
\end{equation}
where $R_c$ is the centrifugal radius, that is the radius at which the radial velocity is equal to the rotation velocity. 

\begin{figure}
\centering
\includegraphics[width=\textwidth,trim={0.2cm 9.3cm 0cm 0.5cm},clip]{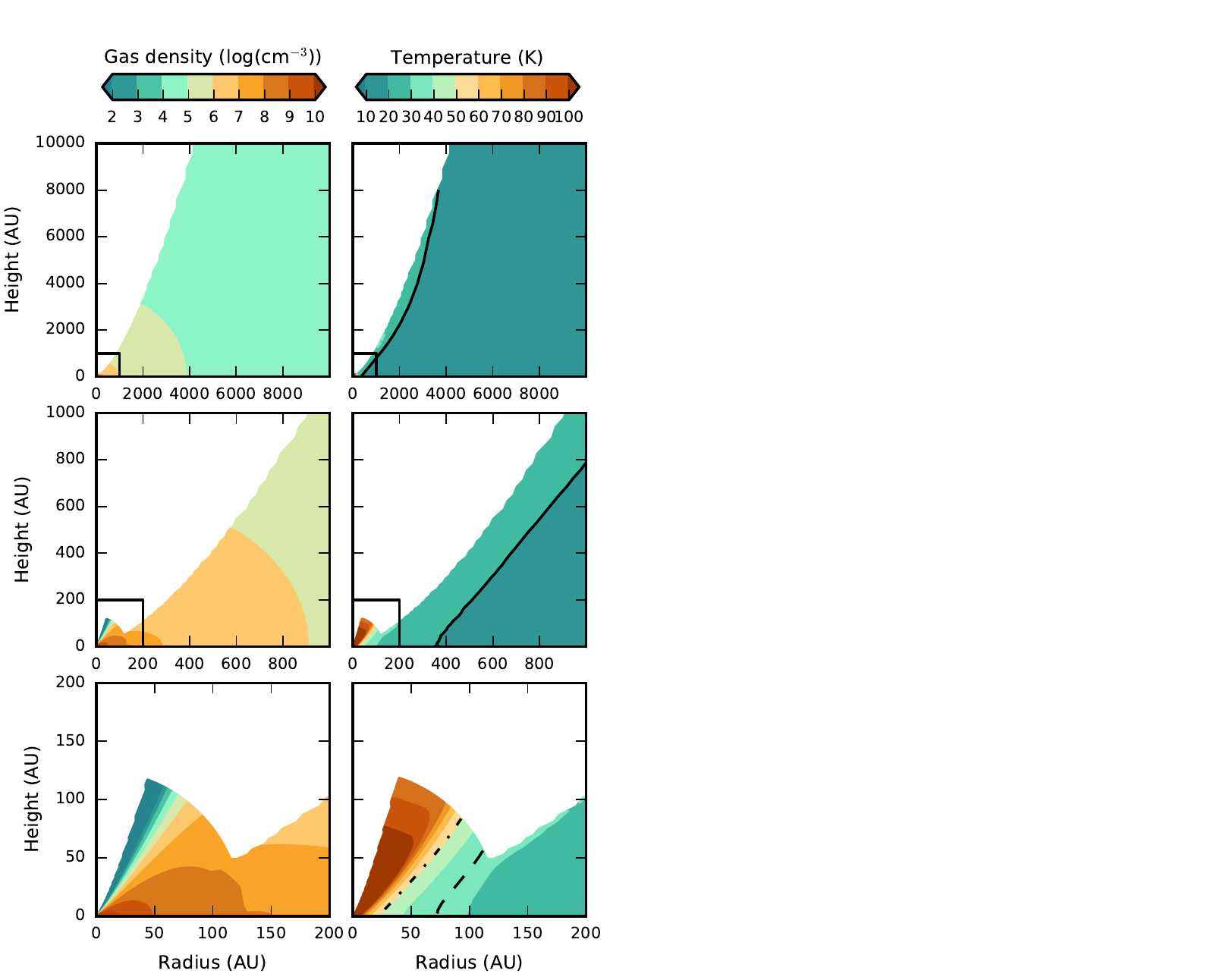}
\caption{Density (\textit{left panels}) and temperature (\textit{right panels}) structure for the disk and envelope of L1527 from the T13 best fit model on a scale of 10,000~AU (\textit{top panels}), 1000~AU (\textit{middple panels}) and 200~AU (\textit{bottom panels}). The black boxes highlight the area shown in the row below. The solid black contours mark the CO snow surface (20 K). The location of the snow surface in the intermediate and cold models are indicated with dashed and dash-dotted contours, respectively.}
\label{fig:2Dstructure}
\end{figure}

The envelope density structure is spherical at large radii, but becomes flattened near the centrifugal radius due to rotation: 
\begin{equation}
\rho_{\rm{env}}(r) = \frac{\dot M_{\rm{env}}}{4\pi(GM_\star r^3)^{1/2}} \left(1 + \frac{\cos\theta}{\cos\theta_0} \right)^{-1/2} \left( \frac{\cos\theta}{\cos\theta_0} + \frac{2\cos^2\theta_0R_c}{r} \right)^{-1},
\end{equation}
where $\dot M_{\rm{env}}$ is the mass infall rate of the envelope onto the disk. Inside $R_c$, $ \rho_{\rm{env}} \propto r^{-1/2} $, while outside $R_c$, $ \rho_{\rm{env}} \propto r^{-3/2} $. The bipolar outflow cavities extend from the central protostar to the outer radius of the envelope. To reproduce the scattered light morphology, a dual-cavity structure was required, with the cavity being more narrow in the inner $\sim$100~AU and wider at larger radii (see Fig.~\ref{fig:2Dstructure}, and \citealt{Tobin2008} for details). Inside the outflow cavity the density is set to zero.

The disk has a Keplerian velocity profile, that is,
\begin{equation} \label{eq:kep}
v_\phi(r) = \sqrt{\frac{GM_\star}{r}}. 
\end{equation}
This means that at radii larger than $R_c$ the velocity structure is given by Eqs~\ref{eq:UCM1}--\ref{eq:UCM2}, while at smaller radii Eq.~\ref{eq:kep} is adopted. A standard flared accretion density structure was assumed for the disk \citep{Shakura1973,Lynden-Bell1974,Hartmann1998}, that is, 
\begin{equation}
\rho_{\rm{disk}}(r) = \rho_0 \left[ 1 - \left( \frac{R_\star}{r} \right)^{1/2} \right] \left( \frac{R_\star}{r}  \right)^\alpha  \exp \left\lbrace  - \frac{1}{2} \left[ \frac{z}{h(r)} \right]^2 \right\rbrace, 
\end{equation}
where the scaleheight $h(r)$ is given by
\begin{equation}
h(r) = h_0 \left( \frac{r}{R_\star} \right)^\beta. 
\end{equation}

\begin{figure}
\centering
\includegraphics[scale=0.8,trim={1cm 1.5cm 0cm 0cm},clip]{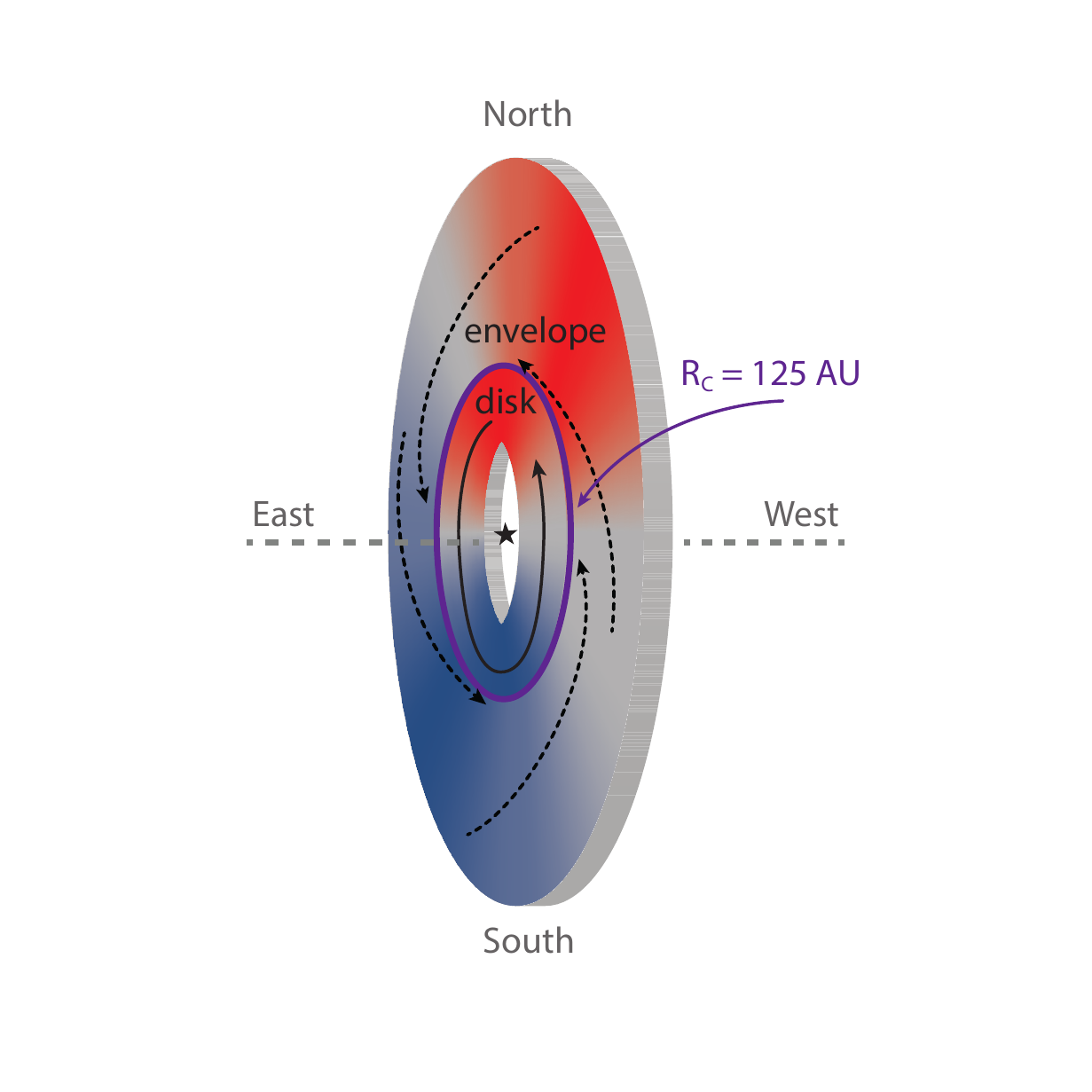}
\caption{Schematic illustration of the disk-envelope model employed here and by T13. The material inside the centrifugal radius, $R_C$, rotates with Keplerian velocity (solid arrow), while the material outside $R_C$ is rotating and infalling (dashed arrows). The blue- and redshifted material is shown in blue and red, respectively. The system is 5$^{\circ}$ inclinded from edge-on, such that the disk and envelope west of the protostar are nearest to the observer.} 
\label{fig:Modelcartoon}
\end{figure}

\begin{figure}
\centering
\includegraphics[width=\textwidth,trim={0.3cm 16cm 0cm 0.5cm},clip]{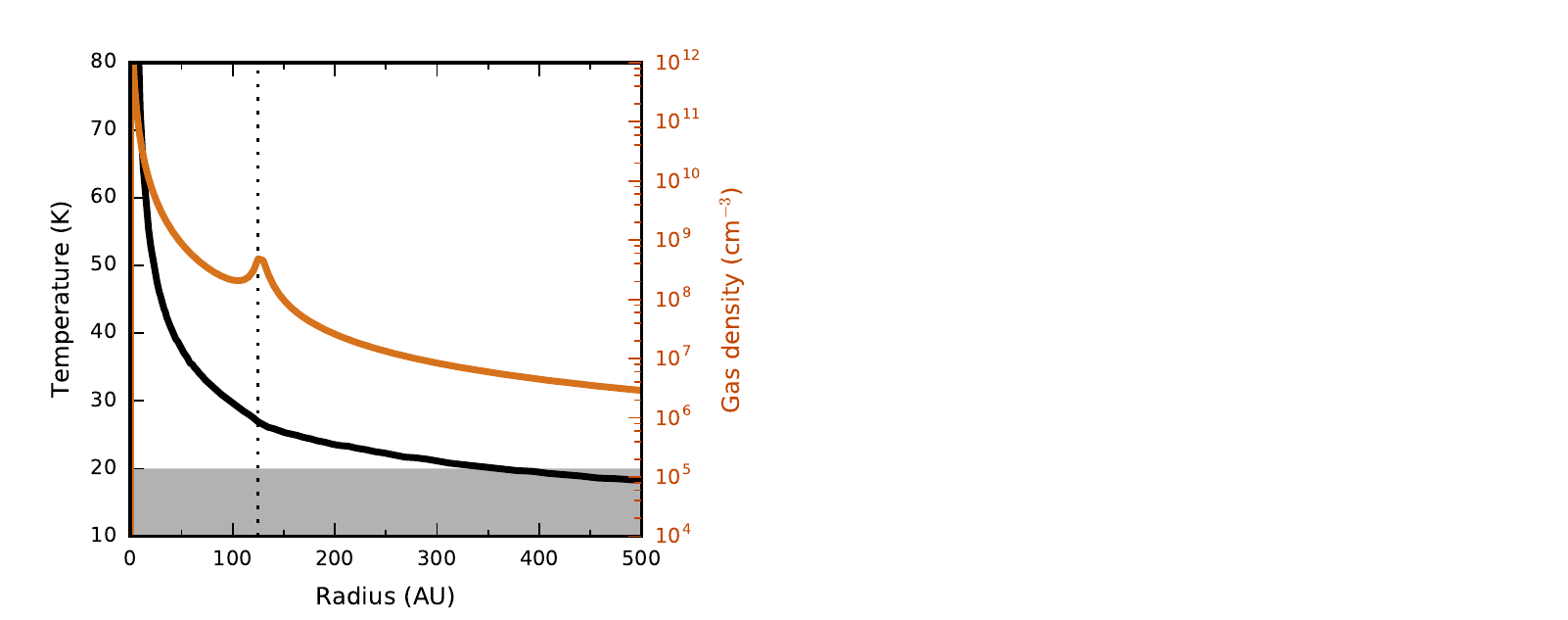}
\caption{Midplane temperature (black) and density (orange) profile from the best fit model from T13 for the disk and inner envelope of L1527. The vertical dotted line marks the disk outer radius and the gray area shows the temperature range (< 20 K) for which CO is expected to be frozen out.}
\label{fig:Midplane_phys}
\end{figure}

To fit the multi-wavelength continuum observations, a grid of models was created by T13 in which the envelope properties were kept constant, while varying the disk properties (radius, flaring, radial density profile, scale height and mass). In addition, the dust spectral index was varied (see T13 for details on the dust properties). A protostellar mass of 0.5$M_\sun$ was adopted, but no parameters directly depend on this value. The mass infall rate was taken to be $1.0\times10^{-5}$ $M_\sun$~yr$^{-1}$. The accretion luminosity was set to $L_{\rm{acc}}$~=~1.75~$L_\sun$ and the protostellar luminosity to 1.0~$L_\sun$. For creating the continuum images, an inclination of 85$^\circ$ was used such that the west side of the disk and envelope is nearest to the observer (see Fig.~\ref{fig:Modelcartoon}). The best-fitting disk model has a disk radius of $R_c$ = 125 AU. The disk was found to be highly flared ($h(r) \propto r^{1.3}$, and $h(100 \rm{AU})$ = 48~AU), and have $\rho_{\rm{disk}} \propto r^{-2.5}$. The 2D density and temperature (calculated from the dust radiative transfer) structure are presented in Fig.~\ref{fig:2Dstructure}, and Fig.~\ref{fig:Midplane_phys} shows the corresponding radial profiles for the midplane.

\section{Free fall velocity}

\begin{figure}
\centering
\includegraphics[width=\textwidth,trim={0.3cm 16cm 0cm 0.1cm},clip]{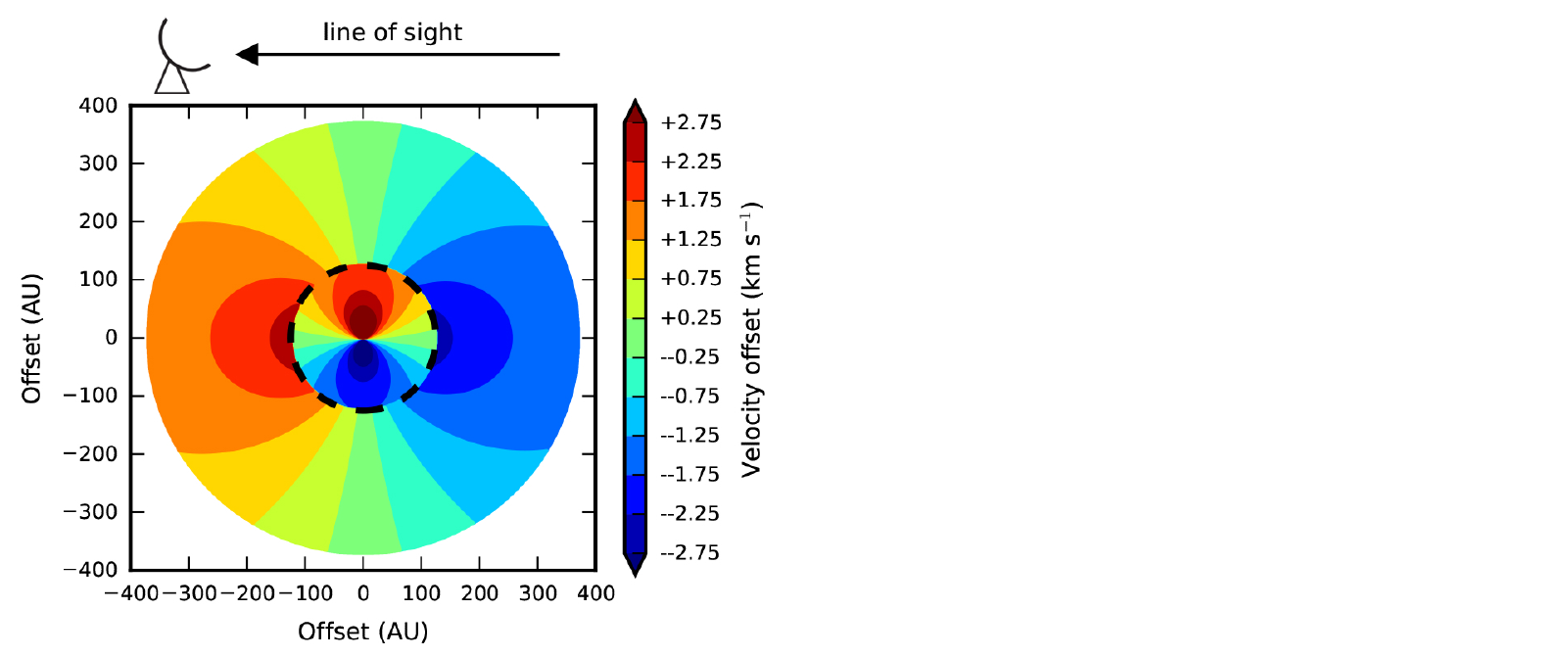}
\caption{As Fig.~\ref{fig:Midplane}, but for a free fall velocity profile in the envelope. }
\label{fig:Midplane_ff}
\end{figure}

\begin{figure*}
\centering
\includegraphics[width=\textwidth]{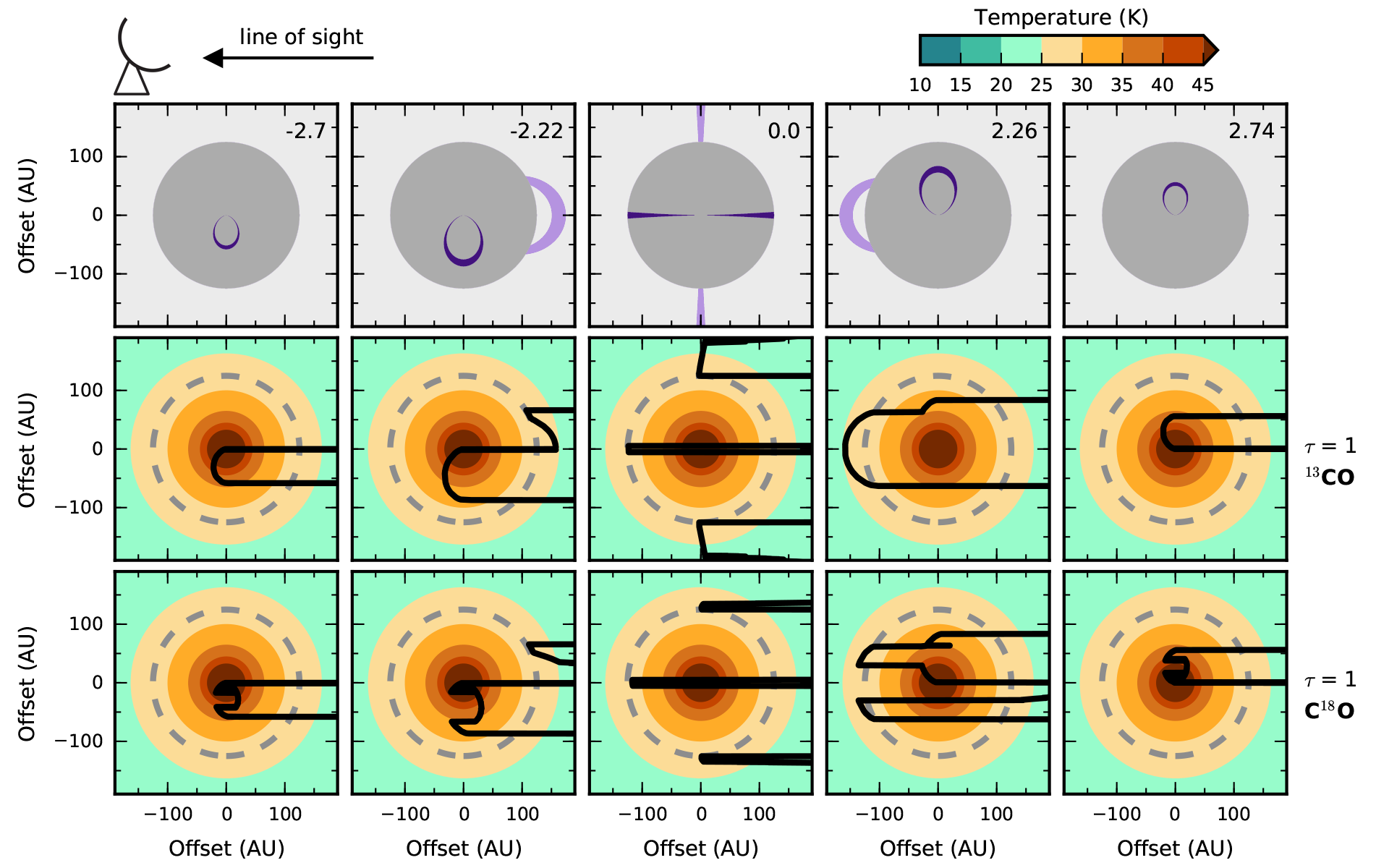}
\caption{As Fig.~\ref{fig:Optdepth}, but for a free fall velocity profile in the envelope, as shown in Fig.~\ref{fig:Midplane_ff}.}
\label{fig:Optdepth_ff}
\end{figure*}

Figure~\ref{fig:Midplane_ff} presents the midplane velocity structure along the line of sight as shown in Fig.~\ref{fig:Midplane}, but for a free fall, 
\begin{equation}
v_r(r) = \sqrt{\frac{2GM_\star}{r}},
\end{equation}
rather than a infalling rotating velocity profile (Eqs.~\ref{eq:UCM1}--\ref{eq:UCM2}) in the envelope. The $\tau = 1$ surface for $^{13}$CO and C$^{18}$O in five representative velocity channels is shown in Fig.~\ref{fig:Optdepth_ff}. 

\end{appendix}

\end{document}